\documentclass[prd,superscriptaddress,amsfonts,amssymb,amsmath,showpacs,twocolumn,nofootinbib]{revtex4-2}
\usepackage{bm}
\usepackage{amsfonts}
\usepackage{latexsym}
\usepackage{graphicx}
\usepackage{amsmath}
\usepackage{palatino}
\usepackage{xcolor} 
\usepackage{mathpazo}
\usepackage{xcolor}
\usepackage{rotating}
\usepackage{adjustbox}
\usepackage{tensor}
\usepackage{textcomp}
\usepackage{float}
\usepackage{booktabs}
\usepackage{dcolumn}
\usepackage[titletoc]{appendix}
\usepackage{booktabs}
\usepackage{multirow}
\usepackage{hyperref}
\hypersetup{colorlinks,citecolor=blue}
\usepackage{amsmath}
\usepackage{xcolor}
\usepackage{orcidlink}
\usepackage{epsfig}
\usepackage{caption}
\usepackage{subcaption}
\usepackage{commath}
\hypersetup{colorlinks,citecolor=blue}
\hypersetup{colorlinks=true,linkcolor=magenta,filecolor=magenta,    urlcolor=blue}

\def\be{\begin{equation}}
\def\ee{\end{equation}}
\def\bea{\begin{eqnarray}}
\def\eea{\end{eqnarray}}

\definecolor{owngreen}{rgb}{0.0, 0.5, 0.0}
\begin{document}

\title{ Exploring the Observational Constraints and Cosmological Dynamics in $f\left(Q,\mathcal{L}_m\right)$ Gravity }

\author{Ratul Mandal\,\orcidlink{0009-0008-6515-5384}}
 \email{ratulmandal2022@gmail.com}
\affiliation{Department of
Mathematics, Indian Institute of Engineering Science and
Technology, Shibpur, Howrah-711 103, India},
\author{Anamika Kotal\,\orcidlink{0009-0009-8426-3206}} 
\email{kotalanamika31@gmail.com}
\affiliation{Department of
Mathematics, Indian Institute of Engineering Science and
Technology, Shibpur, Howrah-711 103, India}
\author{Vipin Kumar Sharma\,\orcidlink{0000-0001-7640-5504
}}  
\affiliation{Department of Physics, School of Chemical Engineering and Physical Sciences,  Lovely Professional University, Phagwara, Punjab, 144411, India}
\affiliation{International Center for High Energy Physics and Applications, Lovely Professional University, Phagwara, Punjab, 144411, India}
\affiliation{Research Center of Astrophysics and Cosmology, Khazar University, Baku, AZ1096, 41 Mehseti Street, Azerbaijan}
\email{vipin.33912@lpu.co.in}
\author{Ujjal Debnath\,\orcidlink{0000-0002-2124-8908}}
 \email{ujjaldebnath@gmail.com}
\affiliation{Department of
Mathematics, Indian Institute of Engineering Science and
Technology, Shibpur, Howrah-711 103, India}

\begin{abstract}
%We explore the cosmological dynamics and observational consistency of the extended gravity framework $f(Q,\mathcal{L}_m)$. In our investigation, 
We explore two scenarios of $f(Q,\mathcal{L}_m)$ gravity: linear and non-linear gravity models. The dynamical system analysis identifies two critical points for each of the proposed linear and nonlinear matter--geometry coupling models. These equilibrium points correspond to distinct phases of cosmic evolution.
%In the linear model, the critical points $P_1$ and $P_2$ describe a matter-dominated scaling solution and a late-time de Sitter accelerating phase, respectively. Similarly, the nonlinear model admits two critical points, $S_1$ and $S_2$, where $S_1$ characterizes an intermediate scaling epoch and $S_2$ represents a stable de Sitter attractor with $\omega_{\rm eff}=-1$ and $q=-1$. 
Depending on the model parameters, the resulting critical points successfully reproduce the observed sequence of cosmic evolution, from a decelerated matter-dominated Universe to the present epoch of accelerated expansion. 
The effective equation of state parameter ($\omega_{\text{eff}}$) and the deceleration parameter ($q$) exhibit smooth transitions from decelerated to accelerated expansion, with transition epochs around $N_{\text{tr}} \approx -0.27$ for linear Model and $N_{\text{tr}} \approx -0.32$ for non-linear Model, consistent with late-time cosmic acceleration. 
Statistical constraints derived from CC+BAO, DESI DR II, and 
Pantheon$^+$ datasets provide best-fit values for the model parameters 
($\alpha, \beta, \gamma, H_0$), showing compatibility with current cosmological 
observations. The analysis employs Akaike (AIC) and Bayesian (BIC) information criteria to evaluate model performance. Our results demonstrate that $f(Q,\mathcal{L}_m)$ gravity provides a viable alternative framework for explaining late-time acceleration, with rich 
dynamical features that merit further exploration in view of upcoming high-precision 
surveys.

%Model comparison using Akaike (AIC) and Bayesian (BIC) information criteria indicates that while the proposed $f(Q,\mathcal{L}_m)$ scenarios yield competitive fits, the standard $\Lambda$CDM model remains statistically favored. Nevertheless, the small deviations observed highlight the potential of matter–geometry couplings to alleviate cosmological tensions, particularly in the determination of the Hubble constant. 

%Employing the dynamical system analysis, we identify critical points and analyze their stability, uncovering late-time attractor solutions that naturally account for the accelerated expansion of the Universe. To assess observational viability, we constrain the model parameters using the latest \textit{Dark Energy Spectroscopic Instrument Data Release 2} (DESI DR2) baryon acoustic oscillation (BAO) measurements, complemented by Type Ia supernova samples (Pantheon$^{+}$, DES-Dovekie, Union3) and CMB shift parameters. Our results demonstrate that both linear and nonlinear matter--geometry coupling scenarios are ..............., with the nonlinear coupling exhibiting ...................... and ...................compared to the linear Model. This study highlights .........................

%We investigated a  $f(Q, L_m)$ cosmological model using recent BAO measurements from DESI DR2, combined with Type Ia supernova samples (Pantheon$^{+}$, DES-Dovekie, and Union3) and CMB shift parameters, to constrain the non-linear model parameters via Markov Chain Monte Carlo analysis. 

\end{abstract}
\maketitle
%\tableofcontents
%%%%%%%%%%%%%%%%%%%%%%%%%%%%%%%%%%%%%%%%%%%%%%%%%
\section{Introduction}
Within the framework of the standard $\Lambda$CDM model based on General Relativity, the Universe is dominated by two primary components: pressureless dark matter (DM, $\sim 26.5\%$) and a hypothetical dark energy (DE, $\sim 68.5\%$) component with negative pressure \cite{aghanim2020planck}. These dark sectors are responsible for both the formation of large-scale structures and the observed late-time accelerated expansion of the Universe, despite the fact that their underlying physical nature remains unknown \cite{Linder:2002et}. The essentially phenomenological character of the $\Lambda$CDM model has therefore motivated the exploration of alternative cosmological scenarios, particularly in light of long standing theoretical issues associated with the cosmological constant—most notably the enormous discrepancy between its observed value (constant in time and behaving as a fluid with fixed energy density) and the predictions of quantum vacuum energy \cite{weinberg1989,Wolf:2023uno,Sahni:1999gb,Padmanabhan:2002ji}.
%Hence, one of the most active research directions over the past decade has focused on interpreting the late-time accelerated expansion of the Universe as an effective extension of the cosmological constant, $\Lambda$. In this context, a wide range of theoretical models have been proposed, which can be broadly classified into two main frameworks for explaining cosmic acceleration: dark energy (DE) and modified gravity (see \cite{Joyce:2016vqv} for a concise review).  A widely studied class of modified gravity theories extends the Einstein-Hilbert action by introducing generic functions of geometric curvature.
Hence, one of the most active research directions over the past decade has focused on interpreting the late-time accelerated expansion of the Universe as an effective extension of the cosmological constant, $\Lambda$. In this context, a wide range of theoretical models have been proposed, which can be broadly classified into two main frameworks for explaining cosmic acceleration: dark energy (DE) and modified gravity (see \cite{Joyce:2016vqv} for a concise review). Within the modified gravity paradigm, several extensions of General Relativity have been explored, including $f(R)$ gravity and its generalizations \cite{DeFelice:2010aj,Sotiriou:2008rp,Amendola,Fazio:2018djb,Nojiri2011,Capozziello2011}. Among these, $f(R,\mathcal{L}_m)$ gravity has attracted particular attention, as it explicitly incorporates non-minimal interactions between the Ricci scalar and the matter Lagrangian, thereby allowing a richer phenomenology arising from matter--geometry couplings \cite{DeFelice:2010aj,Sotiriou:2008rp}. In addition, several alternative gravitational theories have also been proposed, including $f(R,T)$ gravity where $T$ represents trace of energy-momentum tensor ($T_{\mu\nu}$) ~\cite{Harko2011,Moraes2016,Shabani2013} and $f(Q)$ gravity~\cite{Heisenberg2024,Jimenez2018,Jimenez2020,Narawade2022}, in which the gravitational interaction is modified through non-metricity rather than curvature. A further extension, $f(Q,T)$ gravity~\cite{Xu2019,Loo2023}, couples the non-metricity scalar to the trace of the energy--momentum tensor, thereby giving rise to a richer phenomenological structure. In this study, we explore cosmological dynamics within the framework of $f(Q,\mathcal{L}_m)$ gravity, where $Q$ represents the non-metricity scalar and $\mathcal{L}_m$ denotes the matter Lagrangian \cite{Myrzakulov:2024esv}. This theory generalizes non-metricity-based $f(Q)$ gravity by explicitly introducing matter--geometry couplings. The cosmological consequences of $f(Q,\mathcal{L}_m)$ gravity have been investigated in several works, encompassing both its theoretical formulation and observational viability.

%Explicitly, we analyze late-time cosmic acceleration in this modified gravity framework using recent  cosmological observations to place constraints on the model parameters.
In this paper, we extend our analysis by exploring new observational consequences of the $f(Q, L_m)$ scenario using recent DESI DR2 cosmological data sets, complemented by a dynamical systems approach. This framework enables a systematic investigation of the background evolution of the model, its stability properties, and the associated critical points. Moreover, such an analysis provides insight into the qualitative behavior of the cosmological dynamics across different evolutionary epochs and facilitates the identification of stable attractor solutions corresponding to distinct phases of the Universe.

The paper is organized as follows. In Section \ref{sec2}, we introduce the mathematical formulation of the model and derive the corresponding field equations. Section \ref{sec3} discusses the linear and nonlinear cosmological models together with their field equations. Section \ref{sec4} presents the dynamical system analysis. In Section \ref{sec5}, we describe the methodology, the recent observational datasets used in the analysis, and the resulting cosmological parameter constraints in Section \ref{sec6}. Finally, Section \ref{sec7} concludes the paper with a summary of the main results.
%In Section~\ref{sec_4}, we apply the dynamical systems approach to compare the models and analyze their phase-space behavior. Finally, we summarize our results and conclusions in Section~\ref{sec_5}.

%%%%%%%%%%%%%%%%%%%%%%%%%%%%%%%%%%%%%%%%%%%%%%%%
\section{Construction of Gravitational Field Equations in $f\left(Q,\mathcal{L}_m\right)$ Gravity}\label{sec2}
In this section, we present a detailed overview of the underlying mathematical foundation of $f(Q,\mathcal{L}_m)$ gravity theory based on the nonmetricity scalar $Q$ and the matter lagrangian $\mathcal{L}_m$. Then, using the variational principle, we obtain the generalized field equations of the $f(Q,\mathcal{L}_m)$ gravity framework. Additionally, we present the non-conservation of the energy-momentum tensor, emphasizing the consequences arising from the coupling between the matter Lagrangian and the underlying geometric structure.
%%%%%%%%%%%%%%%%%%%%%%%%%%%%%%%%%%%%%%%%%%%%%%%%%%
\subsection{Mathematical Preliminaries}
In the context of Riemannian geometry, the metric tensor $g_{\mu\nu}$ describes the background geometric structure of a space-time, and the affine connection $\Gamma^\lambda_{\mu\nu}$ denotes the parallel transport. For a general metric, the affine connection can be expressed as a combination of the Levi-Civita connection,  contortion, and disformation in the following way \cite{hehl1995metric}
\begin{equation}
    Y^\lambda_{\mu\nu}=\Gamma^\lambda_{\mu\nu}+ K^\lambda_{\mu\nu} +L^\lambda_{\mu \nu} 
\end{equation}
Here $\Gamma^\lambda_{\mu\nu}$ is the Levi-Civita connection, $ K^\lambda_{\mu\nu} $ is contortion and  $L^\lambda_{\mu \nu}$ represents the disformation respectively. During the parallel transport of a vector, the Levi-Civita connection preserves the inner product of tangent vectors, and it is defined as 
\begin{equation}
    \Gamma^\lambda_{\mu\nu}=\frac{1}{2}g^{\lambda \alpha} \left(\partial_{\mu} g_{\mu \alpha}+\partial_{\nu} g_{\alpha \mu} -\partial_{\alpha} g_{\mu \nu} \right)
\end{equation}
The contortion tensor $K^\lambda_{\mu\nu}$ is an antisymmetric tensor with respect to its first two indices and is defined in terms of the torsion tensor $T^\lambda_{\mu\nu}$ in the following manner
\begin{equation}
    K^\lambda_{\mu\nu}=\frac{1}{2}\left(T^\lambda_{\mu\nu}+T_{\mu \;\;\nu}^{\;\lambda}+T_{\nu \;\;\mu}^{\;\lambda}\right)
\end{equation}
The information tensor $L^\lambda_{\mu\nu}$ signifies the contraction or expansion of space time and can be defined in terms of the non-metricity tensor in the following 
\begin{equation}
    L^\lambda_{\mu\nu}=\frac{1}{2}\left(Q^\lambda_{\mu\nu}-Q_{\mu \;\;\nu}^{\;\lambda}-Q_{\nu \;\;\mu}^{\;\lambda}\right)
\end{equation}
Here, the non-metricity tensor $Q_{\lambda\mu\nu}$ measures the variation of the length of a vector along its path during the parallel transport. Mathematically, the non-metricity tensor can be defined as the covariant derivative of the metric tensor $g_{\mu\nu}$ with respect to the affine connection $Y^\lambda_{\mu\nu}$ such that
\begin{equation}
    Q_{\lambda\mu\nu}=\nabla_\lambda g_{\mu\nu}=\partial_\lambda g_{\mu\nu}-Y^\alpha_{\lambda\mu} g_{\alpha\nu}-Y^\alpha_{\lambda\nu}g_{\mu\alpha}
\end{equation}
Additionally, the non-metricity conjugate, also known as the superpotential tensor $P^\lambda_{\mu\nu}$ defined as \cite{xu2019f}
\begin{equation}
    P^\lambda_{\mu\nu}=-\frac{1}{2}L^\lambda_{\mu\nu}+\frac{1}{4}\left(Q^\lambda-\tilde{Q}^\lambda\right)g_{\mu\nu}-\frac{1}{4}\delta^\lambda_{(\mu} Q_{\nu)}
\end{equation}
Here $Q^\lambda=Q^{\lambda}{}_{\mu}{}^{\mu}$ and $\tilde{Q^\lambda}=Q_{\mu}^{\;\lambda\mu}$ are two independept traces called the non-metricity vector. The non-metricity scalar $Q$ encapsulates the deviation of the manifold geometry from isotropy. From a physical perspective, it measures the change of volume experienced by an object undergoing parallel transport. Mathematically, the non-metricity scalar can be constructed by contracting the superpotential tensor with the non-metricity tensor in the following way
\begin{equation}\label{Q}
    Q=-Q_{\lambda\mu\nu}P^{\lambda\mu\nu}
\end{equation}
%%%%%%%%%%%%%%%%%%%%%%%%%%%%%%%%%%%%%%%%%%%%%
\subsection{Gravitational action and Generalized Field Equations}
The Einstien-Hilbert action corresponding to $f\left(Q,L_m\right)$ gravity has the following form
\begin{equation}\label{action}
    \mathcal{S}=\int f\left(Q,\mathcal{L}_m\right)\sqrt{-g} d^4x
\end{equation}
Here $\sqrt{-g}$ is the determinant of metric tensor $g_{\mu\nu}$ i.e $g=\lvert g_{\mu\nu}\rvert$ and $f\left(Q,\mathcal{L}_m\right)$ is an arbitrary analytical function of non-metricity scalar $Q$ and matter Lagrangian $\mathcal{L}_m$ .\\\\
In cosmology, the field equation is extremely important to studying the dynamics of a particular modified gravity or dark energy model, as it describes how the spacetime geometry is related to the matter or energy sector.
In order to derive the field equation corresponding to $f\left(Q,\mathcal{L}_m\right)$ gravity, we use the least action principle, which requires varying the Einstein-Hilbert action \eqref{action} with respect to the metric tensor. For a general metric and arbitrary form of $f\left(Q,\mathcal{L}_m\right)$, the corresponding field equation is obtained as\cite{hazarika2025f}
\begin{eqnarray}\label{gen field eqn}
    &&\frac{2}{\sqrt{-g}}\nabla_\alpha \left(f_Q\sqrt{-g} P^\alpha_{\;\mu\nu}\right)+f_Q \left(P_{\mu\alpha\beta}Q_\nu ^{\;\alpha\beta}-2Q^{\alpha\beta}_{\;\mu} P_{\alpha\beta\nu} \right)+\nonumber\\
    &&\frac{1}{2}fg_{\mu\nu}=\frac{1}{2}f_{\mathcal{L}_m}\left(g_{\mu\nu}\mathcal{L}_m-T_{\mu\nu}\right)
\end{eqnarray}
where $f_Q=\frac{\partial f\left(Q,\mathcal{L}_m\right)}{\partial Q}$ and $f_{\mathcal{L}_m}=\frac{\partial f\left(Q,\mathcal{L}_m\right)}{\partial \mathcal{L}_m}$.\\
Additionally, the energy momentum tensor $T_{\mu\nu}$ related to the matter component is given by
\begin{equation}
    T_{\mu\nu}=-\frac{2}{\sqrt{-g}}\frac{\delta\left(\sqrt{-g}\mathcal{L}_m\right)}{\delta g^{\mu\nu}}=g_{\mu\nu}\mathcal{L}_m-2\frac{\partial\mathcal{L}_m}{\partial g^{\mu\nu}}
\end{equation}
Also, by applying variation to the Einstien-Hilber action with respect to the contention, we get the field equation as
\begin{equation}
    \nabla_\mu\nabla_\nu\left(4\sqrt{-g}f_Q P^{\mu\nu}_{\;\;\alpha}+H_{\alpha}^{\;\;\mu\nu}\right)=0
\end{equation}
Here $H_{\alpha}^{\;\;\mu\nu}$ denotes the hypermomentum density and it is defined as
\begin{equation*}
    H_\alpha ^{\;\;\mu\nu}=\sqrt{-g}f_{\mathcal{L}_m}\frac{\delta \mathcal{L}_m}{\delta Y^{\alpha}_{\;\;\mu\nu}}
\end{equation*}
Like the modified $f\left(R, T\right)$ gravity and Ratsall gravity model, the $f\left(Q,\mathcal{L}_m\right)$ gravity does not satisfy the energy conservation equation. By using the covariant derivative of the field equation, one can obtain the non-conservation equation in the following form
\begin{eqnarray}\label{non conservation}
    D_\mu T^{\mu}_{\;\nu}&=&\frac{1}{f_{\mathcal{L}_m}}\Bigg[\frac{2}{\sqrt{-g}}\nabla_\alpha \nabla_\mu H_\nu ^{\;\alpha\mu}+\nabla_\mu A^\mu_{\;\nu}-\nonumber\\ &&\nabla_\mu \left(\frac{1}{\sqrt{-g}}\nabla_\alpha H_\nu^{\;\;\alpha\mu}\right)\Bigg]\nonumber\\
    &=&B_\nu\neq0
\end{eqnarray}
Due to the presence of the non-conservation tensor $B_\nu$ on the right-hand side of the conservation equation \eqref{non conservation}, the usual conservation of energy momentum is not conserved for the present gravitational framework. The non-conservation tensor $B_\nu$ is a function of dynamical variables such as the non-metricity scalar $Q$ and the matter lagrangian $\mathcal{L}_m$; the nonvanishing existence of $B_\nu$ signifies the interaction between the geometry and matter sector of the particular gravity theory .\\\\
In order to study the cosmological evolution through the dynamical system analysis, we consider the background geometry as a flat, homogeneous, isotropic space-time given by the FLRW metric
\begin{equation}\label{metric}
    ds^2=-dt^2+a^2\left(t\right)\left(dx^2+dy^2+dz^2\right)
\end{equation}
Here $a\left(t\right)$ is the usual scale factor and the Hubble parameter is defined as $H=\frac{\Dot{a\left(t\right)}}{a\left(t\right)}$. For the considered metric \eqref{metric}, the expression of the non-metricity scalar \eqref{Q} can be obtained in terms of the Hubble parameter as $Q=6H^2$.\\
The matter sector for the present cosmological model is considered to be in the form of a perfect fluid given by
\begin{equation}\label{em}
    T_{\mu\nu}=\left(\rho+p\right)u_\mu u_\nu +p g_{\mu\nu}
\end{equation}
Where $\rho$ and $p$ represent the energy density and isotropic pressure, respectively, and $u_\mu$ is the usual four-velocity vector.\\
Under the FLRW metric \eqref{metric} and using the expression of the energy momentum tensor \eqref{em}, we obtain the Friedmann equation from the generalized field equation \eqref{gen field eqn} as
\begin{eqnarray}
    3H^2&=&\frac{1}{4f_Q}\left(f-f_{\mathcal{L}_m}\left(\rho+\mathcal{L}_m\right)\right)\label{1st friedmann}\\
    \Dot{H}+3H^2&=&-\frac{\Dot{f_Q}}{f_Q}H+\frac{1}{4f_Q}\left(f+f_{\mathcal{L}_m}\left(p-\mathcal{L}_m\right)\right)\label{2nd Friedmann}
\end{eqnarray}
Additionally, the generalized non-conservation equation \eqref{non conservation} takes the form
\begin{equation}
    \Dot{\rho}+3H\left(\rho+p\right)=B_\mu u^\mu
\end{equation}
The nonzero right-hand side of the above equation describes the deviation from geodesic motion. The term $B_\mu u^\mu$ is associated with the dissipation of energy. For $B_\mu u^\mu=0$, the system satisfies the general energy conservation law, and for $B_\mu u^\mu \neq0$, the energy transfer process will be dominant.\\\\
One can reformulate the Friedmann equation \eqref{1st friedmann} in its usual form, such as
\begin{eqnarray}
    3H^2&=&\rho_{eff}\label{eq18}\\
    2\Dot{H}+3H^2&=&-p_{eff}\label{eq19}
\end{eqnarray}
Where $\rho_{eff}$ and $p_{eff}$ are the effective density parameter and pressure respectively and defined as 
\begin{eqnarray}
    \rho_{eff}&=&\frac{1}{4f_Q}\left(f-f_{\mathcal{L}_m}\left(\rho+\mathcal{L}_m\right)\right)\\
    p_{eff}&=&2\frac{\Dot{f_Q}}{f_Q}H-\frac{1}{4f_Q}\left(f+f_{\mathcal{L}_m}\left(\rho+2p-\mathcal{L}_m\right)\right)
\end{eqnarray}
Additionally, the Friedmann equation \eqref{eq18} and \eqref{eq19} allow us to write the conservation equation corresponding density and pressure component to its usual form
\begin{equation}\label{eq22}
\Dot{\rho}_{eff}+3H\left(\rho_{eff}+p_{eff}\right)=0
\end{equation}
To study the further cosmological evolution in the present $f\left(Q,\mathcal{L}_m\right)$ gravity model, we define the effective equation of state (Eos) parameter $\omega_{eff}$ and the deceleration parameter $q$ as
\begin{eqnarray}
    \omega_{eff}&=&-1-\frac{2\Dot{H}}{3H^2}\\
    q&=&-1-\frac{\Dot{H}}{H^2}
\end{eqnarray}
The Eos parameter plays an important role in studying cosmological dynamics. Distinct values of the Eos parameter correspond to different phases of the cosmological evolution. Specifically, $\omega_{eff}=1$  corresponds to a stiff fluid–dominated era , $\omega_{eff}=\frac{1}{3},0$ describes a radiation dominated era or matter-dominated epoch respectively, $-1<\omega_{eff}<-\frac
{1}{3}$ characterizes the quintessence regime and $\omega_{eff}=-1$ signifies the De-sitter epoch. In addition, the deceleration parameter is important in studying the nature of cosmological expansion.  $q>0$ represents a decelerated expansion, and $q<0$ describes the accelerating cosmological expansion, which corresponds to the current dark energy-dominated era.
%%%%%%%%%%%%%%%%%%%%%%%%%%%%%%%%%%%%%%%%%%%%%
\section{Cosmological Models Based on $f\left(Q,\mathcal{L}_m\right)$ Gravity}\label{sec3}
In the previous section, we have presented the field equation and other important equations corresponding to the general  $f\left(Q,\mathcal{L}_m\right)$ gravity framework. But in order to construct the dynamical system and study further cosmological implications, it is important to consider a specific form of the function $f$. In this section, we present two distinct cosmological models based on the $f\left(Q,\mathcal{L}_m\right)$ gravity framework by considering two different forms of the function $f$. First, we specify the Lagrangian $\mathcal{L}_m$ of cosmic matter by $\mathcal{L}_m=p$. To maintain the generality of our analysis, we consider the matter component satisfies the usual equation of state given by $p=\left(\gamma-1\right)\rho$, where $p$ and $\rho$ are pressure and energy density, respectively, and $\gamma$ is a constant parameter.
%%%%%%%%%%%%%%%%%%%%%%%%%%%%%%%%%%%%%%%%%%%%%
\subsection{Model-I: $f\left(Q,\mathcal{L}_m\right)=-\alpha Q+2\mathcal{L}_m+\beta$} \label{Model-1}
In the first cosmological model, we consider a linear form of the function $f$ as $f\left(Q,\mathcal{L}_m\right)=-\alpha Q+2\mathcal{L}_m+\beta$, here $\alpha,\beta$ are the constant model parameter and the matter lagrangian $\mathcal{L}_m=p=\left(\gamma-1\right)\rho$. This model was first proposed by Hazarika et al in \cite{hazarika2025f}. Additionally, in \cite{Myrzakulov_2024}, the authors have studied the accelerating expansion under this function. For this specific choice of the functional, the Friedmann equations reduce to
\begin{eqnarray}
    3H^2&=&-\frac{\beta}{2\alpha}+\frac{\rho}{\alpha}\label{eq25}\\
    2\Dot{H}+3H^2&=&-3H^2\left(\gamma-1\right)-\frac{\beta \gamma}{2\alpha}\label{eq26}
\end{eqnarray}
From \eqref{eq25} and \eqref{eq26}, one can identify the expression of effective energy density and pressure in the following form
\begin{eqnarray}
    \rho_{eff}&=&-\frac{\beta}{2\alpha}+\frac{\rho}{\alpha}\label{eq27}\\
    p_{eff}&=&3H^2\left(\gamma-1\right)+\frac{\beta \gamma}{2\alpha}\label{eq28}
\end{eqnarray}
The energy-momentum tensor corresponding to the matter component satisfies the conservation law 
\begin{equation}
    \Dot{\rho}+3H\left(\rho+p\right)=0
\end{equation}
Now, from the first Friedmann equation \eqref{eq25}, we can writedown the expressions of the Hubble parameter $H\left(z\right)$ as a function of redshift $z$ and model parameters as 
\begin{equation}
    H\left(z\right)=\Bigg\{\frac{\left(6H_0^2 \alpha+\beta\right)\left(1+z\right)^{3\gamma}-\beta}{6\alpha}\Bigg\}^{\frac{1}{2}}
\end{equation}
where, $H_0$ is the present day value of Hubble parameter $H\left(z\right)$.
%%%%%%%%%%%%%%%%%%%%%%%%%%%%%%%%%%%%%%%%%%%%%
\subsection{Model-II: $f\left(Q,\mathcal{L}_m\right)=- \alpha Q+\left(2{\mathcal{L}_m}\right)^2+\beta$ } In the second cosmological model, we consider a form of the functional $f$ which replicates the nonlinear effect of the matter lagrangian such that $f\left(Q,\mathcal{L}_m\right)=-\alpha Q+\left(2\mathcal{L}_m\right)^2+\beta$, here $\alpha,\beta$ are constant model parameter. By using this specific form of the function, the Friedmann equations are reduced to the following form
\begin{eqnarray}
    3H^2&=&-\frac{\beta}{2\alpha}-\frac{2}{\alpha}\left(1-\gamma^2\right)\rho^2\\
    2\Dot{H}+3H^2&=&-\frac{\beta}{2\alpha}-\frac{\left(\gamma-1\right)\left(\beta+6\alpha H^2\right)}{2\alpha\left(\gamma+1\right)}
\end{eqnarray}
Therefore, for this particular cosmological model the expressions of the effective energy density and pressure is in the following form
\begin{eqnarray}
    \rho_{eff}&=&-\frac{\beta}{2\alpha}-\frac{2}{\alpha}\left(1-\gamma^2\right)\rho^2\\
    p_{eff}&=&\frac{\beta}{2\alpha}+\frac{\left(\gamma-1\right)\left(\beta+6\alpha H^2\right)}{2\alpha\left(\gamma+1\right)}
\end{eqnarray}
By using the above expressions of effective energy density and pressure, we obtain the energy balance equation in the following form
\begin{equation}
    \Dot{\rho}+\frac{3\gamma}{\gamma+1}H\rho=0
\end{equation}
From the first Friedmann equation, the expressions of the Hubble parameter $H(z)$ can be obtained as
\begin{equation}
    H\left(z\right)=\Bigg\{\frac{\left(6H_0^2 \alpha+\beta\right)\left(1+z\right)^{\frac{6\gamma}{1+\gamma}}-\beta}{6\alpha}\Bigg\}^{\frac{1}{2}}
\end{equation}
%%%%%%%%%%%%%%%%%%%%%%%%%%%%%%%%%%%%%%%%%%%%%%%%%%

\section{Dynamical System Analysis}\label{sec4}
In this section, we have presented a detailed dynamical system analysis corresponding to $f\left(Q,\mathcal{L}_m\right)$ gravity for Model-I and Model-II. The dynamical system analysis provides a rich mathematical framework to study the dynamics of any cosmological model. A detailed review of dynamical system analysis applied in cosmological models is presented in the following literature \cite{bahamonde2018dynamical,coley1999dynamical}. For a cosmological framework 
%%%%%%%%%%%%%%%%%%%%%%%%%%%%%%%%%%%%%%%%%%

%%%%%%%%%%%%%%%%%%%%%%%%%%%%%%%%%%%%%%%%%%%%%%%%%
\subsection{ Formulation of the Dynamical System for Model-I}
In order to construct the dynamical system corresponding to Model-I, first, we reformulate the corresponding Friedmann equation \eqref{eq25} in the following form
\begin{equation}
    1=-\frac{\beta}{6\alpha H^2}+\frac{\rho}{3\alpha H^2}
\end{equation}
The dynamical variable corresponding to Model I is 
\begin{equation}\label{eq38}
    x=\frac{1}{H^2},y=\frac{\rho}{3H^2}
\end{equation}
The variable introduced in \eqref{eq38}, also known as the expansion normalized variable. In this context, one may use a different set of variables to formulate the dynamical system, and in this scenario, the resulting dynamical system will be different, but the main dynamical feature of the system will be identical. However, differentiating the dynamical variable with respect to the e-folding parameter $N=\log a(t)$, we get the dynamical system as 
\begin{eqnarray}
    \frac{dx}{dN}&=&-2x\frac{\Dot{H}}{H^2}\\
    \frac{dy}{dN}&=&\frac{\Dot{\rho}}{3H^3}-2y\frac{\Dot{H}}{H^2}
\end{eqnarray}
Due to the presence of $\frac{\Dot{H}}{H^2}$ and $\frac{\Dot{\rho}}{3H^3}$ in the above dynamical system, the dynamical system is not closed. In order to close the dynamical system, it is important to determine these expressions in terms of the dynamical variable.\\
Now using the expression of the effective density and pressure from \eqref{eq27} and \eqref{eq28} in \eqref{eq22}, we obtain
\begin{equation}\label{eq41}
    \frac{\Dot{\rho}}{3H^3}=\frac{\beta}{2}x-3y-3\alpha\left(\gamma-1\right)-\frac{\beta \gamma}{2}x
\end{equation}
Also, the second Friedmann equation\eqref{eq26} can be rewritten in terms of the dynamical variable as
\begin{equation}\label{eq42}
    \frac{\Dot{H}}{H^2}=-\frac{3}{2}\left(\gamma+\frac{\beta \gamma x}{6\alpha}\right)
\end{equation}
Using the expressions from \eqref{eq41} and \eqref{eq42} , the final form of the autonomous dynamical system is 
\begin{eqnarray}
    \frac{dx}{dN}&=&3x\left(\gamma+\frac{\beta \gamma x}{6\alpha}\right)\label{eq43}\\
    \frac{dy}{dN}&=&\frac{\beta}{2}x-3y-3\alpha\left(\gamma-1\right)-\frac{\beta \gamma}{2}x+\nonumber\\
    &&3y\left(\gamma+\frac{\beta \gamma x}{6\alpha}\right)\label{eq44}
\end{eqnarray}
Additionally, using equation \eqref{eq42} we can obtain the expression of effective Eos parameter $\omega_{eff}$ and the deceleration parameter $q$ in terms of the dynamical variable as
\begin{eqnarray}
    \omega_{eff}&=&-1+\left(\gamma+\frac{\beta \gamma x}{6\alpha}\right)\\
    q&=&-1+\frac{3}{2}\left(\gamma+\frac{\beta \gamma x}{6\alpha}\right)
\end{eqnarray}
\subsection{Critical Points and Stability Analysis}
To continue the further analysis, our next step is to find the critical point corresponding to the dynamical system \eqref{eq43} -\eqref{eq44}. In order to obtain the critical points, we solve the right-hand side of \eqref{eq43}-\eqref{eq44} with zero. We have found a total of two critical points corresponding to Model-I, listed as $P_1$ and $P_2$ respectively. In the table \ref{Critical points table}, we presented the critical points $P_1$ and $P_2$ respectively with their viable existence condition and corresponding value of Eos and deceleration parameter. The stability property of each critical point is studied through the linear stability theory by evaluating the Jacobian matrix's eigenvalues. To study the physical properties, we associate each critical point with some particular cosmological epoch by evaluating the value of the Eos parameter $\omega_{eff}$ and the deceleration parameter $q$. A detailed study of two critical points is presented in the next paragraph. 
\begin{table}
\centering
\begin{tabular}{cccccc}
\hline
&&&&&\\
         Critical point &$x$&$y$&$\omega_{eff}$&$q$&Existence  \\
			&&&&&Condition\\
			\hline
			\hline
            &&&&&\\
$P_1$&0&$\alpha$&$-1+\gamma$&$\frac{3 \gamma }{2}-1$&Always\\
&&&&&\\
			$P_2$&-$\frac{6\alpha}{\beta}$&$0$&$-1$&$-1$& $\beta\neq0$ \\
            &&&&&\\
            \hline
\end{tabular}
\caption{Critical points along with their existence condition for Model-I}
\label{Critical points table}
\end{table}
\begin{itemize}
    \item \textbf{Critical Point $P_1$:} The cosmological solution related to the first critical point $P_1$ is exist throughout the phase space. Since the value of the $x$ coordinate is zero, this family of critical points represents the effect of the matter Lagrangian. The value of the effective Eos parameter corresponding to $P_1$ is $\omega_{eff}=-1+\gamma$. Different values of $\gamma$ lead us to different cosmological epochs. For example, $\gamma=0$, lead us to the de-sitter era where the dark energy component behave as a cosmological constant, $0<\gamma<\frac{2}{3}$ implies $-1<\omega_{eff}-\frac{1}{3}$ representing a quintessence like behavior and $\gamma=1$ can representsa matter dominated era with $\omega_{eff}=0$. Similarly, the value of the deceleration parameter corresponding to $P_1$ is $q=-1+\frac{3\gamma}{2}$. Therefore, for an accelerated cosmological expansion, i.e, $q<0$, the value of $\gamma$ must satisfy the condition $\gamma<\frac{2}{3}$ and for $q>\frac{2}{3}$ this cosmological solution represents a decelerated phase of our universe. In order to determine the stability nature, we obtain the eigenvalue of the Jacobian matrix at $P_1$ as
    \begin{equation*}
     \left\lbrace 3 (\gamma -1),3 \gamma\right\rbrace 
    \end{equation*}
    Both eigenvalues are nonzero and functions of $\gamma$ imply the critical point is hyperbolic and therefore, linear stability theory can be applied to determine the stability. For $\gamma<0$, both eigenvalues are negative, and according to the linear stability theory, this critical point is stable or an attractor until $\gamma<0$. For $0<\gamma<1$, the eigenvalues are of opposite sign, and this critical point will exhibit a saddle-like behavior, and for $\gamma>1$, this critical point is unstable due to all positive eigenvalues.
\end{itemize}
\begin{itemize}
    \item \textbf{Critical Point $P_2$:}The second critical point $P_2$ exist in the phase space unless $\beta\neq0$. Due to the vanishing $y$ coordinate that is related to the matter density parameters, this critical point always represents a completely dark energy-dominated cosmological era. The value of the effective Eos parameter for $P_2$ is obtained as $\omega_{eff}=-1$, representing the de-sitter epoch where the dark energy component behaves as a cosmological constant. Similarly, the value of the deceleration parameter at $P_2$ is $q=-1$. The negative value of the deceleration parameter indicates the cosmological accelerated expansion. The eigenvalues of the Jacobian matrix corresponding to the critical point $P_2$ are
    \begin{equation*}
        \left\lbrace-3,-3\gamma\right\rbrace
    \end{equation*}
    Since both eigenvalues are nonzero, $P_2$ is a hyperbolic critical point, and the linear stability theory can be applied to determine the stability behavior. If $\gamma>0$, then the second eigenvalue will be negative and the critical point will exhibit a stable nature, and for $\gamma<0$, the second eigenvalue will be positive and the critical point exhibits a saddle-type behavior.
\end{itemize}
%%%%%%%%%%%%%%%%%%%%%%%%%%%%%%%%%%%%%%%%%%%%%%%%%%%%%%%%%%%%%
\begin{figure}
    \centering
    \includegraphics[width=1\linewidth]{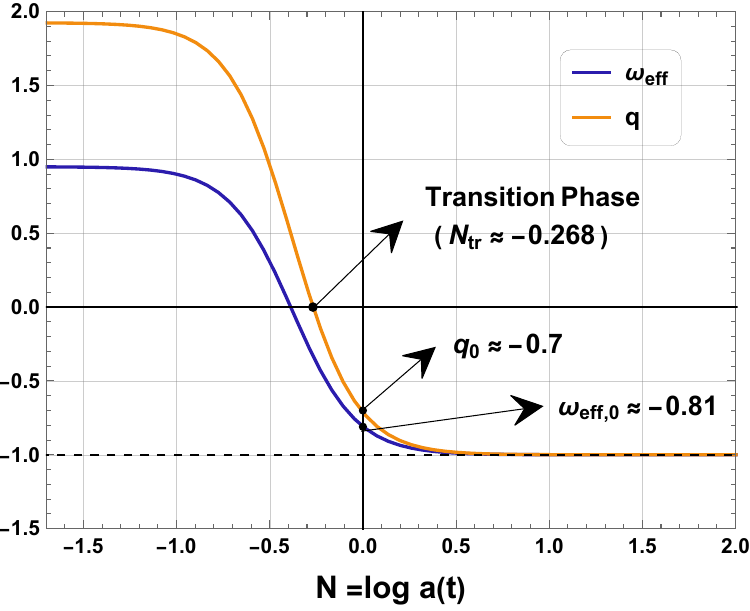}
    \caption{Evolution of the effective Eos parameter $\omega_{eff}$ and the deceleration parameter $q$ corresponding to Model-I. The vertical line $N=0$ represents the present epoch, while $N>0$ and $N<0$ indicates the present and past epoch, respectively}
    \label{fig:placeholder}
\end{figure}
%%%%%%%%%%%%%%%%%%%%%%%%%%%%%%%%%%
By numerically integrating the dynamical system using fine tuned initial condition, we have tracked the evolution of effective Eos parameter $\omega_{eff}$ and the deceleration parameter $q$ respectively. In Figure \ref{fig:placeholder}, we have presented the evolution of $\omega_{eff}$ and $q$ corresponding to Model I. The evolution of these cosmographic parameters can depicts the main cosmological epochs associated with the critical points. As it can be seen from Figure \ref{fig:placeholder}, the effective Eos parameter strat evolving from the prematter-dominated epoch ($\omega_{eff}>0$) and then passes through the radition dominated era ($\omega_{eff}=\frac{1}{3}$) and matter dominated era ($\omega_{eff}=0$), it succesfully transit through the quintessense epoch ($-1<\omega<-\frac{1}{3}$) and finally converves to the de-sitter era $\omega_{eff}=-1$ in the late time. The present value of the effective Eos parameter is obtained as $\omega_{eff,0}\approx-0.81$, showing that our Universe currently is in the quintessence stage. Similarly, the evolution of the deceleration parameter $q$ presented in Figure \ref{fig:placeholder}, portrays a transition from decelerated to acclerated era in recent time at $N_{tr}=-0.268$. The present value of the deceleration parameter we obtain is $q_{0}=-0.7$. This negative value of $q$ represents the current accelerated expansion that the universe is going through.
%%%%%%%%%%%%%%%%%%%%%%%
\begin{figure*}
    \centering
    \begin{subfigure}{0.32\textwidth}
        \centering
        \includegraphics[width=\linewidth]{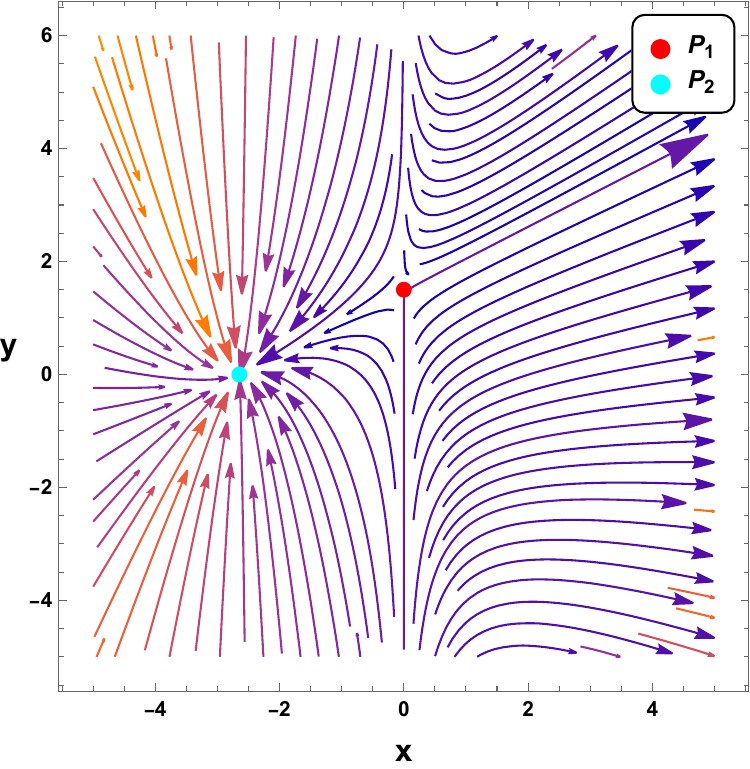}
        \caption{Trajectories in Phase-Space for the parameter $\gamma \in(0,1)$}
    \end{subfigure}
    \hfill
    \begin{subfigure}{0.32\textwidth}
        \centering
        \includegraphics[width=\linewidth]{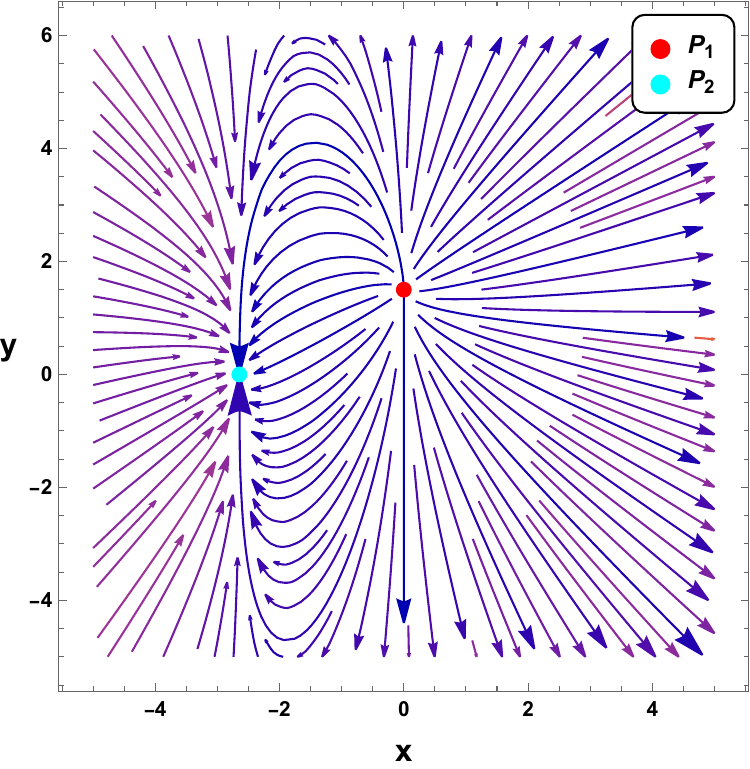}
        \caption{Trajectories in Phase-Space for the parameter $\gamma>1$  }
    \end{subfigure}
    \hfill
    \begin{subfigure}{0.32\textwidth}
        \centering
        \includegraphics[width=\linewidth]{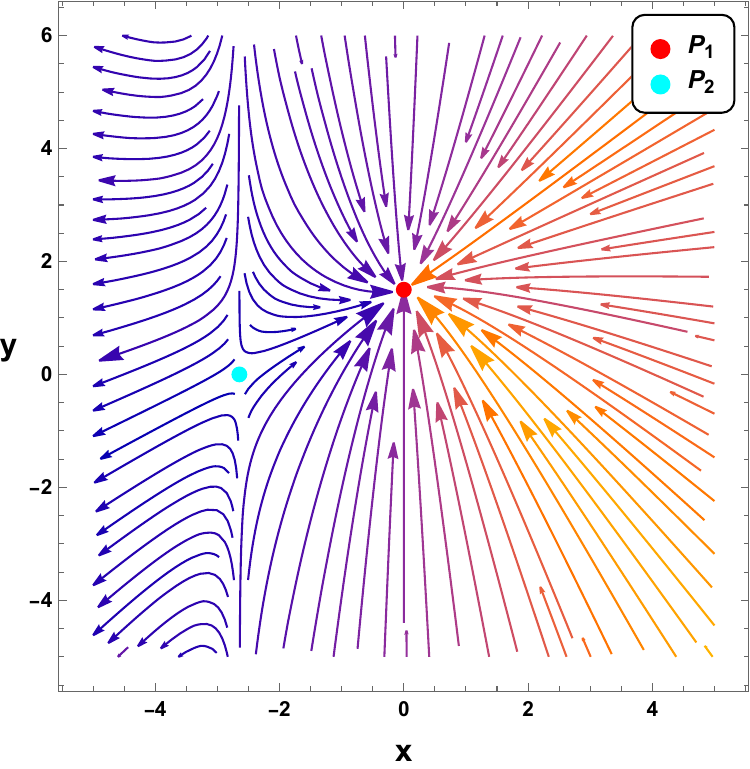}
        \caption{Trajectories in Phase-Space for the parameter $\gamma<0$}
    \end{subfigure}
    \caption{Phase space trajectories around critical points of Model-I}
\end{figure*}
%%%%%%%%%%%%%%%%%%%%%%%%%%%%%%%%%%%%%%%%%

%%%%%%%%%%%%%%%%%%%%%%%%%%%%%%%%%%%%%%%%%

\subsection{ Formulation of the Dynamical System for Model-II }
To construct the autonomous dynamical system corresponding to the functional $f(Q,\mathcal{L}_m)=-\alpha Q+(2\mathcal{L}_m)^2+\beta$, where $\alpha,\beta$ are arbitrary model parameter and $Q$ is nonmetricity term and $\mathcal{L}_m$ is Lagrangian density , we rewrite the corresponding Friedmann equation in the following form
\begin{eqnarray}
    1=-\frac{\beta}{6\alpha H^2}-\frac{2}{\alpha}(1-\gamma^2)\frac{\rho^2}{3H^2}
\end{eqnarray}
To formulate the dynamical system, we have considered the following dynamical variable 
\begin{eqnarray}
    x=\frac{1}{H^2},y=\frac{\rho^2}{3H^2}
\end{eqnarray}
Now, differentiating the dynamical variable with respect to the "e-folding" parameter $N$, we obtain the dynamical system as follows
\begin{eqnarray}
    \frac{dx}{dN}&=&-\frac{6\gamma}{\gamma+1}x-2x\frac{\dot{H}}{H^2}\label{eq49}\\
    \frac{dy}{dN}&=&-\frac{2}{3}\frac{\dot{H}}{H^4}\label{eq50}
\end{eqnarray}
Due to the pressence of the term $\frac{\dot{H}}{H^2}$ and $\frac{\dot{H}}{H^4}$, the dynamical system is not closed. To close the dynamical system, We calculate the value of these term from the  corresponding accleration equation as follows
\begin{eqnarray}
    \frac{\dot{H}}{H^2}&=&-\frac{3\beta y}{4\alpha}-\frac{3(\gamma-1)(\beta y+2\alpha)}{4\alpha(\gamma+1)}-\frac{3}{2}\label{eq51}\\
    \frac{\dot{H}}{H^4}&=&3y\left(-\frac{3\beta y}{4\alpha}-\frac{3(\gamma-1)(\beta y+2\alpha)}{4\alpha(\gamma+1)}-\frac{3}{2}\right)\label{eq52}
\end{eqnarray}
Now using the expression presented in \eqref{eq51} and \eqref{eq52} to the dynamical system \eqref{eq49}-\eqref{eq50}, we get the dynamical system in its desired form as
\begin{eqnarray}
    \frac{dx}{dN}&=&-\frac{6 \gamma  x}{\gamma +1}-2 x \left(-\frac{3 (\gamma -1) (2 \alpha +\beta  y)}{4 \alpha  (\gamma +1)}-\frac{3 \beta  y}{4 \alpha }-\frac{3}{2}\right)\nonumber\\
    &&\label{eq53}\\
    \frac{dy}{dN}&=&-\frac{2}{3} 3 y \left(-\frac{3 (\gamma -1) (2 \alpha +\beta  y)}{4 \alpha  (\gamma +1)}-\frac{3 \beta  y}{4 \alpha }-\frac{3}{2}\right)\label{eq54}
\end{eqnarray}
Using \eqref{eq51}, the expressions for the effective Eos parameter $\omega_{eff}$ and the deceleration parameter $q$ corresponding to Model-II is obtained as
\begin{eqnarray}
    \omega_{eff}&=&\frac{\alpha  (\gamma -1)+\beta  \gamma  y}{\alpha  (\gamma +1)}\\
    q&=&\frac{\alpha  (4 \gamma -2)+3 \beta  \gamma  y}{2 \alpha  (\gamma +1)}
\end{eqnarray}
%%%%%%%%%%%%%%%%%%%%%%%%%%%%%%%%%%%%%%%%%%%%%%%%%%%%%%%%%%%%%%%%%%%%%%%%%%%%%%%%%%%%%%%%%%%%%%%%
\begin{table}\label{table2cp}
\centering
\begin{tabular}{cccccc}
\hline
&&&&&\\
         Critical point &$x$&$y$&$\omega_{eff}$&$q$&Existence  \\
			&&&&&Condition\\
			\hline
			\hline
            &&&&&\\
$S_1$&Any&$0$&$\frac{-1+\gamma}{1+\gamma}$&$\frac{2 \gamma -1}{\gamma +1}$&Always\\
&&&&&\\
			$S_2$&$0$&$-\frac{2\alpha}{\beta}$&$-1$&$-1$& $\beta\neq0$ \\
            &&&&&\\
            \hline
\end{tabular}
\caption{Critical points along with their existence condition for Model-II}
\label{Critical points table 2}
\end{table}
%%%%%%%%%%%%%%%%%%%%%%%%%%%%%%%%%%%%%%%%%%%%%%%%%%
\subsection{Critical Points and Their Stability Analysis}
To obtain the critical point, we solve the right-hand side of the system of equations \eqref{eq53}-\eqref{eq54}. We got two physically viable critical point namely, $S_1$ and $S_2$. The coordinate of these critical point in the $(x,y)$ phase space along their existense criteria and associated background cosmological parameters like effective Eos parameter $\omega_{eff}$ and the deceleration parameter $q$ are presented in Table\ref{Critical points table 2}.
\begin{itemize}
    \item \textbf{Critical point $S_1$:}The cosmological solution corresponding to the critical point $S_1$ charecterized the effect of geometric sector only. The value of the effective Eos parameter at this family of critical point is obtained as $\omega_{eff}=\frac{-1+\gamma}{1+\gamma}$. One can recover the matter dominated era  corresponding to $\omega_{eff}=0 $ through this critical point by considering $\gamma=1$. Simmilarly,  for $0<\gamma<\frac{1}{2}$, these particularsolution exhibts the quintessence epoch corresponding to $-1<\omega_{eff}<-\frac{1}{3}$. Moreover, a phantom era can be achived through the solution by considering $\gamma\in(-1,0)$. The eigenvalues of the Jacobian matrix corresponding to this critical point is obtained as 
    \begin{equation*}
     \left\lbrace -\frac{6 \gamma }{\gamma +1},-\frac{6 \gamma }{\gamma +1} \right\rbrace 
    \end{equation*}
    The critical points are hyperbolic in nature since both the eigenvalues are nonzero functions of $\gamma$ and one can use linear stability theory to determine the stability. Under the parameter range $\gamma\in(-\infty,-1)\cup(0,\infty)$, both eigenvalues are negative and hence under this range, this family of critical points will exhibit a stable node-like behavior. On the other hand, for $\gamma\in(-1,0)$ both the eigenvalues are positive and hence the corresponding eigenvectors are divergent and the critical point will exhibit an unstable behavior. 
\end{itemize}
\begin{itemize}
    \item \textbf{Critical point $S_2$:} The second family of the critical point always exists in the phase space unless $\beta\neq0$. The value of the effective Eos parameter corresponding to $S_2$ is $\omega_{eff}=-1$. This constant negative value of the effective Eos parameter implies that the cosmological solution associated with this critical point will represent the de sitter epoch of evolution, where the dark energy sector mimics the cosmological constant like behavor. Also the constant negative value of the deceleration parametr $q=-1$ represent a acclerated expansion of the universe. The eigenvalues of the Jacobian matrix corresponding to $S_2$  are 
    \begin{equation*}
     \left\lbrace 0,\frac{6 \gamma }{\gamma +1} \right\rbrace 
    \end{equation*}
 Due to the existence of one vanishing eigenvalue, this critical point is normally hyperbolic, and its stability depends on the sign of the nonvanishing eigenvalue. Thus, within the region $\gamma\in(-1,0)$ this critical point exhibt stabe behavior and for $\gamma\in(\infty,-1)\cup(0,\infty)$ the critical point will be unstable.  
\end{itemize}
%%%%%%%%%%%%%%%%%%%%%%%%%%%%%%
\begin{figure}
    \centering
\includegraphics[width=1\linewidth]{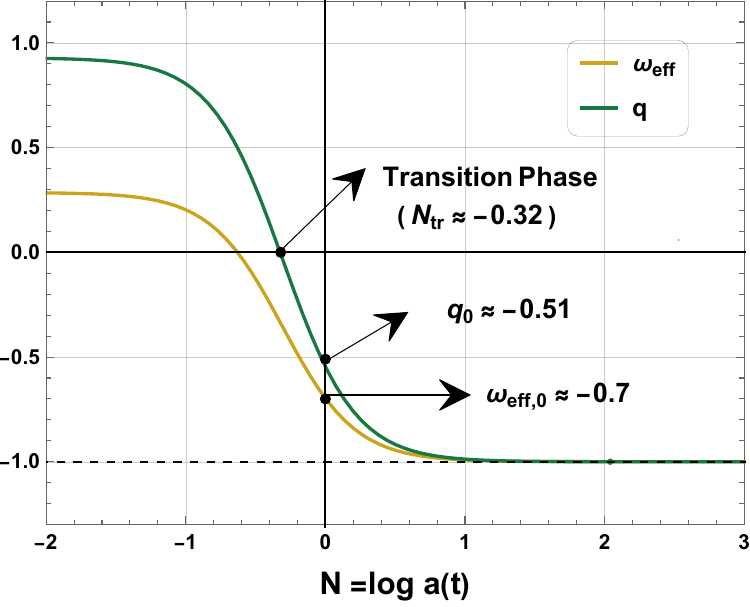}
    \caption{Evolution of the effective Eos parameter $\omega_{eff}$ and the deceleration parameter $q$ corresponding to Model-II. The vertical line $N=0$ represents the present epoch, while $N>0$ and $N<0$ indicates the present and past epoch, respectively}
    \label{fig:eos2}
\end{figure}

%%%%%%%%%%%%%%%%%%%%%%%%%%%%%%%%
 Figure \ref{fig:eos2} illustrates the evolution of effective Eos parameter $\omega_{eff}$ and the deceleration parameter $q$ corresponding to Model II . The evolution of the effective Eos parameter begins with prematter-dominated epoch ($\omega_{eff}>0$),  passes through the radition ($\omega_{eff}=\frac{1}{3}$) and matter dominated era ($\omega_{eff}=0$), and then  transition through the quintessense regime ($-1<\omega<-\frac{1}{3}$) and finally converves to the de-sitter era $\omega_{eff}=-1$ in the late time. The present value $\omega_{eff,0}\approx-0.7$, showing that our Universe currently is in the quintessence stage. The evolution of the deceleration parameter $q$  illustrates an expected transition from decelerated to acclerated era in recent time at $N_{tr}=-0.32$. Its present value $q_{0}=-0.51$ incicate the current acclerated expansion and compitable with the observational evidences.
%%%%%%%%%%%%%%%%%%%%%%%%%%%%%%%%
\begin{figure*}
    \centering
    \begin{subfigure}{0.32\textwidth}
        \centering
        \includegraphics[width=\linewidth]{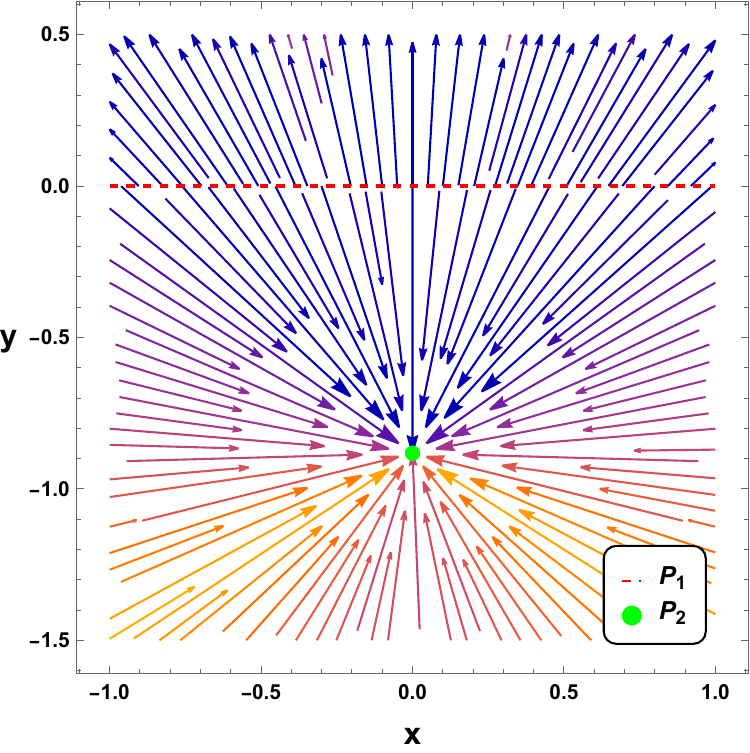}
        \caption{Trajectories in Phase-Space for the parameter $\gamma \in(-\infty,-1)$}
    \end{subfigure}
    \hfill
    \begin{subfigure}{0.32\textwidth}
        \centering
        \includegraphics[width=\linewidth]{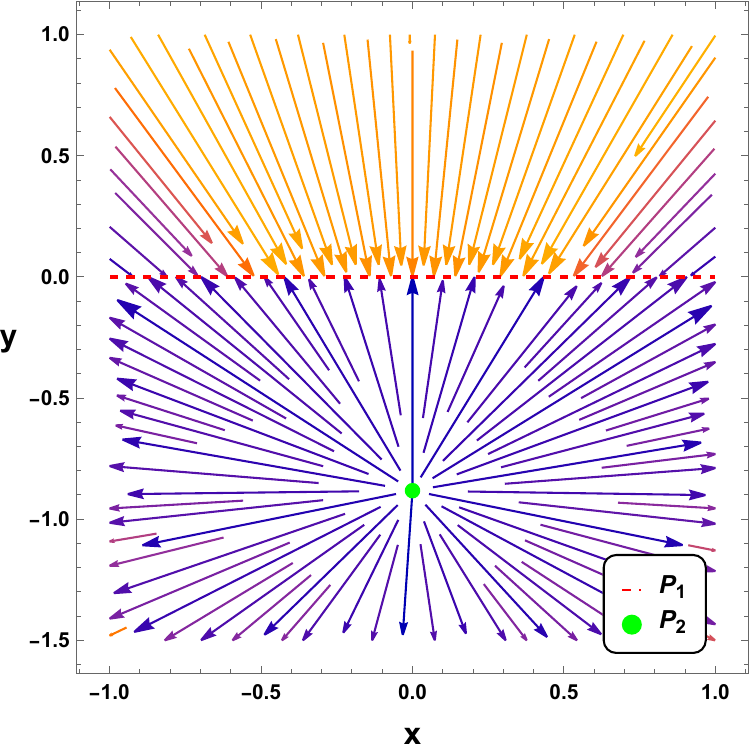}
        \caption{Trajectories in Phase-Space for the parameter $\gamma\in(-1,0)$  }
    \end{subfigure}
    \hfill
    \begin{subfigure}{0.32\textwidth}
        \centering
        \includegraphics[width=\linewidth]{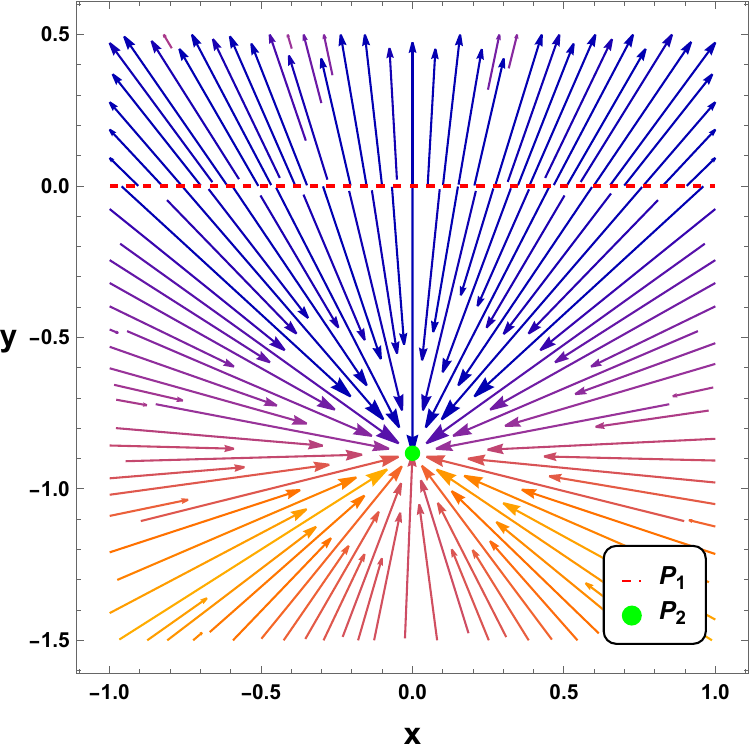}
        \caption{Trajectories in Phase-Space for the parameter $\gamma\in(0,\infty)$}
    \end{subfigure}
    \caption{Phase space trajectories around critical points of Model-II}
\end{figure*}

%%%%%%%%%%%%%%%%%%%%%%%%%%%%%%%
%%%%%%%%%%%%%%%%%%%%%%%%%%%%%%%%%%%%%%%%%%%%%%%%

%%%%%%%%%%%%%%%%%%%%%%%%%%%%%%%%%%%%%%%%%%%%%%%%%%%%%%%%%%%%%%%%%%%

\section{Methodology and Datasets}\label{sec5}
 We constrain the free parameters of our modified gravity model ( Model I and Model II) using three independent observational datasets. These datasets collectively probe both the geometric and dynamical aspects of cosmic evolution across a wide range of redshifts, ensuring robust and complementary constraints on the model parameter space.

 \subsection{Observational Hubble Data (OHD)  from cosmic chronometers(CC)}
 The Cosmic Chronometer method provides a direct, model-independent measurement of the Hubble parameter $H(z)$ by exploiting the differential age evolution of passively evolving galaxies \cite{Magana:2017nfs, Sharov:2018yvz, Moresco:2018xdr}. The underlying principle follows from the definition:
\begin{equation}
    H(z) = -\frac{1}{1+z}\frac{dz}{dt}
    \label{eq:hubble_cc}
\end{equation}
By measuring the redshift difference $dz$ and the associated age interval $dt$ inferred from pairs of massive, early-type galaxies that originated during a similar epoch but are observed at marginally different redshifts, one obtains $H(z)$ without assuming any cosmological model. This model independence is a key advantage: CC measurements serve as an unbiased, cosmology-agnostic anchor for the expansion history of the Universe. For the present analysis, we consider a dataset comprising $N_{\rm CC} = 32$ measurements of the Hubble parameter $H(z)$, covering the redshift interval $0.07 \leq z \leq 1.965$, drawn from 
\cite{Moresco2012, Moresco2015, Moresco2016, Zhang2014, Ratsimbazafy2017}. 
\subsection{Baryon Acoustic Oscillations (BAO)}

Baryon Acoustic Oscillations arise from acoustic pressure waves that propagated through the tightly coupled baryon-photon plasma in the early Universe, prior to recombination. At the redshift $z_d$, corresponding to the drag epoch, the baryonic matter ceased to remain kinetically coupled to the photon fluid; these oscillations became frozen into the matter distribution, imprinting a characteristic length scale known as the \textit{sound horizon}:
\begin{equation}
    r_d = \int_{z_d}^{\infty} \frac{c_s(z)}{H(z)}\, dz
    \label{eq:sound_horizon}
\end{equation}
where the sound speed of the baryon-photon fluid is given by:
\begin{equation}
    c_s(z) = \frac{c}{\sqrt{3\left(1 + \dfrac{3\Omega_b}{4\Omega_\gamma(1+z)}\right)}}
    \label{eq:sound_speed}
\end{equation}
with $\Omega_b$ and $\Omega_\gamma$ being the present-day baryon and photon density parameters, respectively. This scale, $r_d \approx 147\,\rm Mpc$, serves as a powerful \textit{standard ruler} that can be detected in the two-point correlation function of the galaxy distribution at different redshifts, providing robust constraints on both the Hubble parameter $H(z)$ and the angular diameter distance $D_A(z)$.
Depending on the orientation of the BAO measurement with respect to the line of sight, different distance combinations are probed. The transverse comoving distance is:
\begin{equation}
    D_M(z) = \frac{c}{H_0} \int_0^z \frac{dz'}{E(z')}
    \label{eq:dm}
\end{equation}
The Hubble distance is:
\begin{equation}
    D_H(z) = \frac{c}{H(z)}
    \label{eq:dh}
\end{equation}
and the spherically averaged distance, commonly reported by spectroscopic surveys, is:
\begin{equation}
    D_V(z) = \left[z\, D_M^2(z)\, D_H(z)\right]^{1/3}
    \label{eq:dv}
\end{equation}
These quantities are typically reported as dimensionless ratios 
$D_M/r_d$, $D_H/r_d$, and $D_V/r_d$, where $r_d$ is the sound horizon at the drag epoch. In this work, we use a compilation of BAO measurements drawn from several major spectroscopic galaxy surveys, including the 6dF Galaxy Survey \cite{Beutler2011}, the SDSS Main Galaxy Sample \cite{Ross2015}, BOSS DR12 \cite{Alam2017}, eBOSS DR16 \cite{Alam2021}, and DESI DR1 \cite{DESI2024}, spanning the redshift range 
$0.106 \leq z \leq 2.334$. In this analysis, we combine a total of \textbf{57 data points} - comprising the CC and BAO compilations described above, and treat them as a single joint dataset. The additive joint $\chi^2$:
\begin{equation}
    \chi^2_{\rm CC+BAO} = \chi^2_{\rm CC} + \chi^2_{\rm BAO}
    \label{eq:chi2_ccbao}
\end{equation}
Their combination, therefore, yields significantly tighter and more robust posteriors than either dataset alone.
\subsection{DESI DR II}
To constrain the cosmological model, we employ the most recent Baryon Acoustic Oscillation (BAO) observations obtained by the Dark Energy Spectroscopic Instrument (DESI) as part of Data Release~2 (DR2). These observations offer precise measurements of cosmological distance scales across the redshift range $0.295 \leq z \leq 2.33$. The DESI DR2 compilation traces the large-scale structure of the Universe through several galaxy and quasar populations, including Bright Galaxy Survey (BGS) galaxies, Luminous Red Galaxies (LRGs), Emission Line Galaxies (ELGs), and quasars\cite{Wang:2015tua, Beutler:2011hx}. The combination of these tracers enhances the statistical precision of the BAO observations and enables tighter constraints on the expansion history of the Universe \cite{Silva:2025hxw, cheng2026beyond, lee2026shape}.

The BAO information is expressed through compressed distance measurements normalized by the sound horizon at the drag epoch, $r_d$. In particular, the DESI DR2 dataset provides measurements of the transverse comoving distance, $D_M/r_d$, the volume-averaged distance, $D_V/r_d$, and the radial Hubble distance, $D_H/r_d$. In this work, the DESI DR2 BAO data are incorporated into the analysis using the covariance matrices and likelihood prescription provided by the DESI Collaboration \cite{Rodrigues:2025tfg, deSouza:2025rhv}.

\subsection{Type Ia Supernovae: Pantheon+SH0ES}

Type Ia supernovae constitute a well-established class of \textit{standardizable candles} for cosmological studies, after 
applying empirical corrections for light-curve, their peak luminosities provide precise measurements of the luminosity distance $D_L(z)$ \cite{SupernovaCosmologyProject:2008ojh, Pan-STARRS1:2017jku}. The observed quantity is the \textit{distance modulus}:
\begin{equation}
    \mu_{\rm obs} = m_B - M_B
    \label{eq:mu_obs}
\end{equation}
Here, $m_B$ denotes the observed peak apparent magnitude in the rest-frame $B$-band, while $M_B$ represents the corresponding absolute magnitude. The theoretical distance modulus is given by:
\begin{equation}
    \mu_{\rm th}(z,\, \boldsymbol{\theta}) = 
    5\log_{10}\left(\frac{D_L(z,\, \boldsymbol{\theta})}{10\,\rm pc}\right)
    \label{eq:mu_th}
\end{equation}
where the luminosity distance is:
\begin{equation}
    D_L(z,\, \boldsymbol{\theta}) = c\,(1+z)
    \int_0^z \frac{dz'}{H(z',\, \boldsymbol{\theta})}
    \label{eq:dl}
\end{equation}

The \textbf{Pantheon$^+$} sample \cite{Scolnic2022, Brout2022} comprises 1701 observed light curves associated with 1550 distinct SNe~Ia, distributed over the redshift range $0.001 \leq z \leq 2.26$, drawn from 18 different photometric surveys. The \textbf{SH0ES} Cepheid calibration \cite{Riess2022} anchors the absolute magnitude $M_B$ using 42 SN~Ia host galaxies observed with the Hubble Space Telescope, breaking the well-known $H_0$--$M_B$ 
degeneracy and directly constraining the local Hubble constant:
\begin{equation}
    H_0^{\rm SH0ES} = 73.04 \pm 1.04 \;\rm km\,s^{-1}\,Mpc^{-1}
    \label{eq:H0_shoes}
\end{equation}

The $\chi^2$ statistic is constructed using the full $1701 \times 1701$ 
covariance matrix $\mathbf{C}_{\rm SN}$, which encodes both statistical 
and systematic uncertainties — including photometric calibration, Milky 
Way dust extinction, peculiar velocity corrections, and intrinsic scatter:
\begin{equation}
    \chi^2_{\rm SN} = \Delta\boldsymbol{\mu}^{T}\, 
    \mathbf{C}_{\rm SN}^{-1}\, 
    \Delta\boldsymbol{\mu}
    \label{eq:chi2_sn}
\end{equation}
where the residual vector is:
\begin{equation}
    \Delta\mu_i = \mu_{\rm obs,i} - 
    \mu_{\rm th}(z_i,\, \boldsymbol{\theta}), 
    \qquad i = 1,\ldots, 1701
    \label{eq:delta_mu}
\end{equation}
 The corresponding Gaussian likelihood is:
\begin{equation}
    \mathcal{L}_{\rm SN}(\boldsymbol{\theta}) \propto 
    \exp\left(-\frac{\chi^2_{\rm SN}}{2}\right)
    \label{eq:like_sn}
\end{equation}

We perform Bayesian parameter estimation using the Markov Chain Monte Carlo (MCMC) technique along with the \texttt{emcee} package \cite{foreman2013emcee}, which employs an affine-invariant ensemble sampling algorithm. The posterior probability distribution of the parameter set
\[
\boldsymbol{\theta}=\{H_0, \alpha, \beta, \gamma\},
\]
where $\alpha$, $\beta$, and $\gamma$ represent the model-dependent parameters, is obtained from Bayes' theorem as \cite{Schwarz:1978tpv, Spiegelhalter:2002yvw}

\[
\mathcal{P}(\boldsymbol{\theta}\,|\,\mathbf{d})
\propto
\mathcal{L}(\mathbf{d}\,|\,\boldsymbol{\theta})\,
\pi(\boldsymbol{\theta}),
\]

where $\mathcal{L}(\mathbf{d}\,|\,\boldsymbol{\theta})$ is the likelihood function associated with the observational data $\mathbf{d}$ and $\pi(\boldsymbol{\theta})$ denotes the prior distribution. Throughout the analysis, we adopt uniform (flat) priors over physically motivated intervals for all free parameters, the corresponding ranges being listed in Table~\ref{tab:priors}.

The MCMC analysis is carried out using $N_{\rm walkers}$ ensemble walkers evolved for $N_{\rm steps}$ sampling steps. To ensure reliable parameter estimation, an initial burn-in phase is removed from the chains. The marginalized one-dimensional and two-dimensional posterior distributions are visualized using the \texttt{GetDist} package \cite{lewis2025getdist}. From these distributions, we obtain the best-fit values and the corresponding marginalized constraints. The two-dimensional contour plots are presented with both $1\sigma$ (68.3\%) and $2\sigma$ (95.4\%) confidence regions, providing a comprehensive visualization of the parameter uncertainties and correlations \cite{kotal2026fractal, kotal2026parameter}.
To evaluate the statistical performance of the proposed model and to compare it with the standard $\Lambda$CDM cosmology, we calculate the Akaike Information Criterion (AIC) and Bayesian Information Criterion (BIC) \cite{liddle2007information, burnham2004multimodel}, defined respectively as

\begin{equation}
{\rm AIC} = -2\ln \mathcal{L}_{\rm max} + 2k,
\end{equation}

\begin{equation}
{\rm BIC} = -2\ln \mathcal{L}_{\rm max} + k\ln N,
\end{equation}

where $\mathcal{L}_{\rm max}$ denotes the maximum likelihood, $k$ is the number of free model parameters, and $N$ represents the total number of observational data points \cite{Akaike:1974vps, schwarz1978estimating}. These information criteria provide a quantitative measure of the balance between the goodness of fit and model complexity. Models with lower AIC and BIC values are generally preferred. To facilitate comparison between competing models, we consider the differences

\begin{equation}
\Delta {\rm AIC} = {\rm AIC}_{\rm model} - {\rm AIC}_{\rm min},
\end{equation}

\begin{equation}
\Delta {\rm BIC} = {\rm BIC}_{\rm model} - {\rm BIC}_{\rm min},
\end{equation}

where $ AIC_{ min}$ and $ BIC_ {min}$ denote the minimum AIC and BIC values obtained among the models considered. Conventionally, $\Delta {\rm AIC}$ or $\Delta {\rm BIC} < 2$ indicates substantial support for a model relative to the preferred model, values in the range $2$-$6$ suggest moderate evidence against the model, while values exceeding $6$ are generally interpreted as strong evidence against it \cite{jeffreys1998theory}.
\begin{table}[ht]
\centering
\caption{Uniform prior ranges adopted for the MCMC analysis in the two Models. The same priors are used for all observational datasets (CC+BAO, Pantheon+SH0ES, and DESI DR2).}
\label{tab:priors}
\renewcommand{\arraystretch}{1.2}
\begin{tabular}{lccccc}
\hline
\textbf{Model} & $H_0$ & $\alpha$ & $\beta$ & $\gamma$ \\
\hline
Model I  & $[65, 75]$
         & $[0.1, 2.5]$
         & $[0.1,7]$
         & $[0.1,0.9]$ \\

Model II & $[65,75]$
         & $[0.5,2.5]$
         & $[1,9]$
         & $[0.01,0.9]$ \\
\hline
\end{tabular}
\end{table}

%%%%%%%%%%%%%%%%%%%%%%%%%%%%%%%%%%%%%%%%%%%%%%%%%%
\section{RESULTS }\label{sec6}
A comparison of the six best-fit results obtained from the three observational datasets and the two models is presented in the table. \ref{Observational Value}. The Hubble constant, $H_0$, exhibits negligible sensitivity to the chosen functional form of the model but shows a strong dependence on the dataset employed. Specifically, the OHD+BAO dataset consistently yields $H_0 \simeq 68.4$ in both Model~1 and Model~2, whereas the Pantheon$^{+}$ and DESI DR~II datasets favor significantly larger values in the range $H_0 \simeq 72.2$-$72.9$, irrespective of the Model considered. Since the OHD+BAO compilation is anchored by cosmic chronometer measurements and BAO standard rulers calibrated through early-Universe physics, while Pantheon$^{+}$ and DESI are more sensitive to late-time distance indicators, this discrepancy reflects the well-known Hubble tension within the $f(Q, L_M)$ framework. Consequently, the model reproduces the preferred expansion rate of each observational probe rather than providing a mechanism to reconcile the tension, and this behavior remains robust across both model Models.\\

\begin{table}
\centering
\caption{Best-fit parameters obtained from CC+BAO, DESI DR II, and Pantheon$^+$ datasets}
\renewcommand{\arraystretch}{1.6}
\footnotesize
\begin{tabular}{|c|c|c|c|c|}
\hline
\textbf{Model} & \textbf{Parameter} & \textbf{CC+BAO} & \textbf{DESI DR II} & \textbf{Pantheon$^+$} \\
\hline
\multirow{4}{*}{Model 1} & $H_{0}$ & $68.35_{-1.0}^{+0.82}$ & $72.85_{-0.45}^{+0.45}$ &  $72.24_{-0.23}^{+0.18}$ \\
\cline{2-5}
                         & $\alpha$ & $1.30_{-0.72}^{+0.60}$ & $1.21_{-0.54}^{+0.69}$ & $1.40_{-0.64}^{+0.64}$\\
\cline{2-5}
                         & $\beta$ & $3.7_{-1.9}^{+1.9}$ & $2.8_{-2.1}^{+1.7}$ & $4.4_{-2.0}^{+2.4}$ \\
\cline{2-5}
                         & $\gamma$ & $0.752_{-0.015}^{+0.015}$ & $0.6901_{-0.0078}^{+0.0078}$ & $0.502_{-0.021}^{+0.021}$ \\
\hline
\multirow{4}{*}{Model 2} & $H_{0}$ & $68.43_{-0.76}^{+0.85}$ & $72.94_{-0.38}^{+0.38}$ & $72.24_{-0.18}^{+0.18}$ \\
\cline{2-5}
                         & $\alpha$ & $1.40_{-0.55}^{+0.55}$ & $1.38_{-0.52}^{+0.52}$ & $1.49_{-0.76}^{+0.62}$ \\
\cline{2-5}
                         & $\beta$ & $4.9_{-2.2}^{+2.2}$ & $3.7_{-2.5}^{+1.7}$ & $5.3_{-2.5}^{+2.9}$ \\
\cline{2-5}
                         & $\gamma$ & $0.598_{-0.018}^{+0.014}$ & $0.5252_{-0.012}^{+0.0095}$ &  $0.337_{-0.014}^{+0.014}$ \\
\hline
\end{tabular}
\label{Observational Value}
\end{table}

On the other hand, the coupling parameters $\gamma$ and $\beta$ display sensitivity to both the observational dataset and the adopted model. For a fixed Model, the parameter $\gamma$ decreases systematically as the dataset changes from the broad-redshift OHD+BAO compilation to the intermediate-redshift DESI BAO measurements and finally to the low-redshift Pantheon$^{+}$ supernova sample. This monotonic behavior suggests that the effective non-metricity matter coupling strength may be influenced by the redshift range probed by the observations, rather than representing a strictly redshift-independent quantity. Such a trend may indicate a scale or epoch-dependent manifestation of the coupling, although a definitive conclusion would require further investigation with additional observational probes. Within a given dataset, the transition from Model~1 to Model~2 consistently leads to lower values of $\gamma$ and higher values of $\beta$, indicating a nontrivial correlation between these parameters. The correlation appears to be more pronounced for the Pantheon$^{+}$ dataset, compared to the OHD+BAO dataset. This difference implies that the broader redshift coverage of OHD+BAO provides additional constraining power that partially alleviates the parameter degeneracy, whereas the predominantly low-redshift Pantheon$^{+}$ sample leaves the correlation more significant. The $\beta$-$\gamma$ degeneracy emerges as an intrinsic feature of the $f(Q, L_M)$ coupling sector, which could be further constrained through future observations spanning a wider redshift range and complementary growth probes such as $f\sigma_8$.

\begin{table}[H]
\centering
\caption{AIC, BIC, and their corresponding differences ($\Delta$AIC and $\Delta$BIC) for the proposed model and the $\Lambda$CDM model using different observational datasets.}
\label{tab:model_metrics_comparison}
\begin{tabular}{|l|c|c|c|c|}
\hline
\textbf{Model} & \textbf{AIC} & \textbf{BIC} & $\mathbf{\Delta}$\textbf{AIC} & $\mathbf{\Delta}$\textbf{BIC} \\
\hline

\multicolumn{5}{|c|}{\textbf{CC+BAO Dataset}} \\
\hline
Model I & 45.23 & 53.40 & 8.99 & 13.08 \\
Model II & 44.22 & 52.43 & 7.98 & 12.11 \\
$\Lambda$CDM & 36.24 & 40.32 & 0.00 & 0.00 \\
\hline

\multicolumn{5}{|c|}{\textbf{DESI DR II Dataset}} \\
\hline
Model I & 20.08 & 24.97 & 4.96 & 8.88 \\
Model II & 21.10 & 26.02 & 5.98 & 9.93 \\
$\Lambda$CDM & 15.12 & 16.09 & 0.00 & 0.00 \\
\hline

\multicolumn{5}{|c|}{\textbf{Pantheon$^{+}$ Dataset}} \\
\hline
Model I & 1702.81 & 1712.57 & 3.98 & 2.86 \\
Model II & 1704.08 & 1713.84 & 5.25 & 4.13 \\
$\Lambda$CDM & 1698.83 & 1709.71 & 0.00 & 0.00 \\
\hline

\end{tabular}
\end{table}

\begin{figure*}
    \centering
    \begin{subfigure}{0.48\textwidth}
        \includegraphics[width=\textwidth]{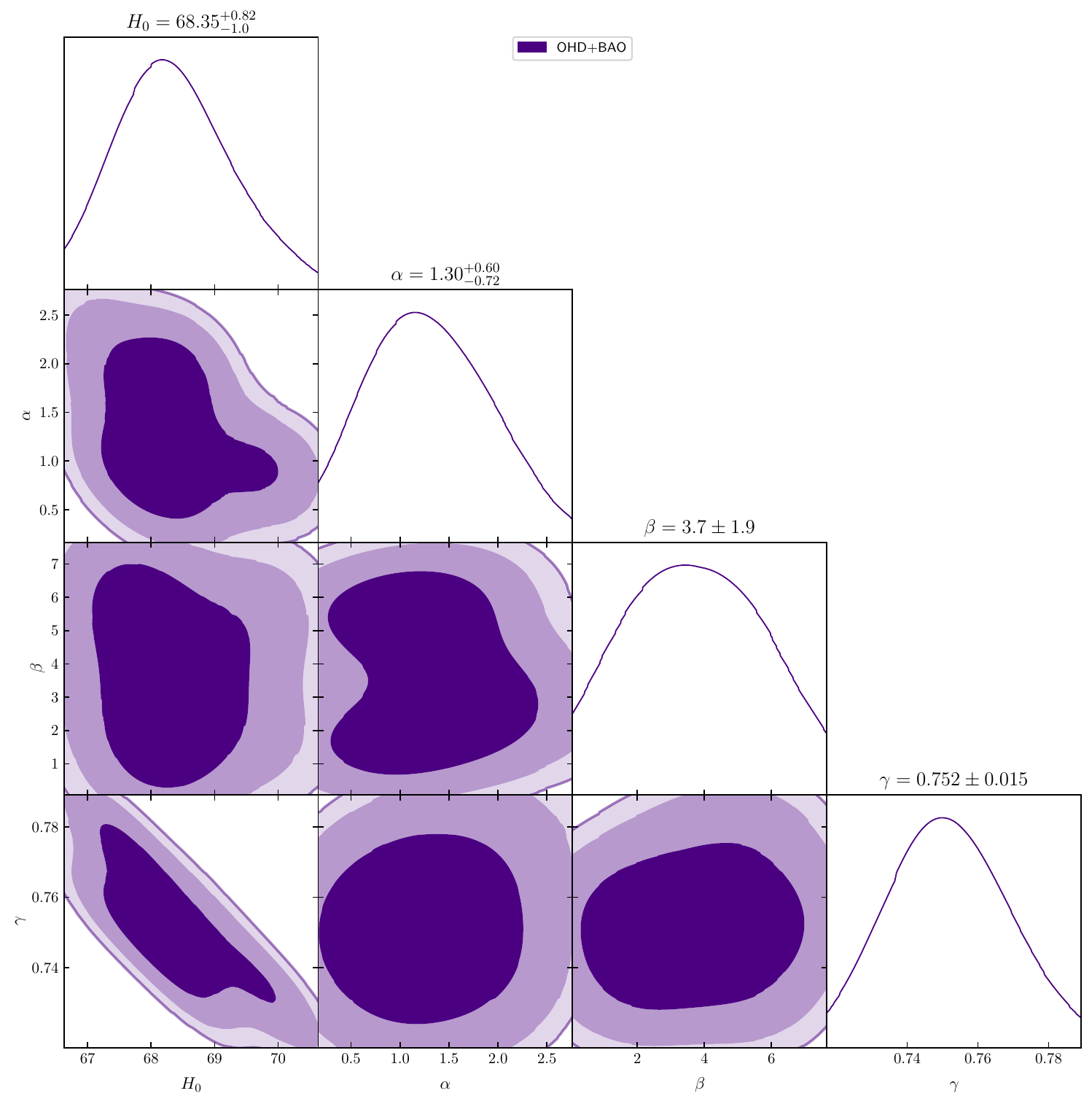}
        \caption{}
    \end{subfigure}
    \begin{subfigure}{0.48\textwidth}
        \includegraphics[width=\textwidth]{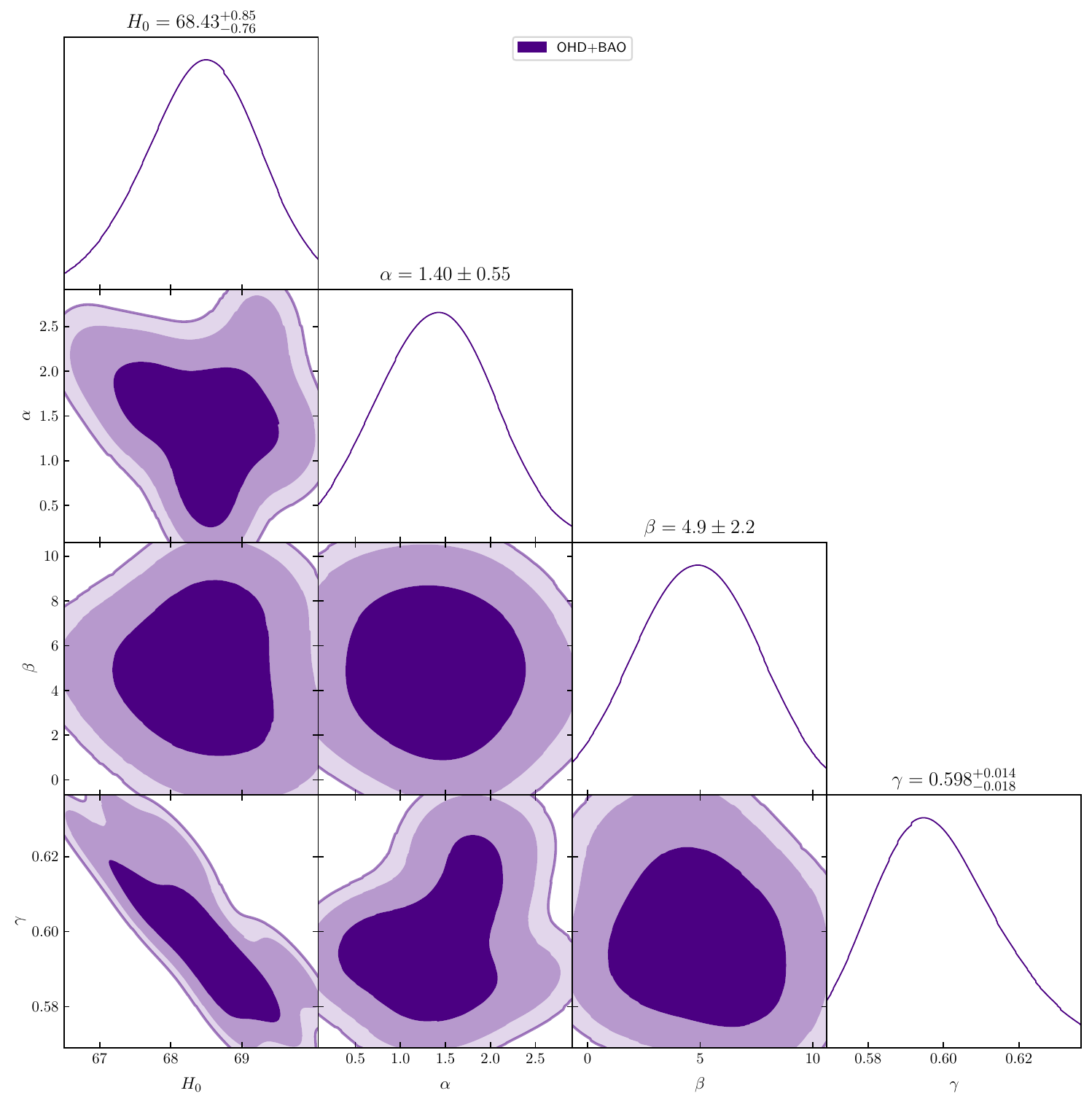}
        \caption{}
    \end{subfigure}
    \caption{Posterior distribution of $f(Q,\mathcal{L}_m)$ gravity for (a)Model-I, (b) Model-II models parameters at $1\sigma$ and $2\sigma$ confidence levels for 57 OHD+BAO data .}
    \label{cc}
\end{figure*}
\begin{figure*}
    \centering
    \begin{subfigure}{0.48\textwidth}
        \includegraphics[width=\textwidth]{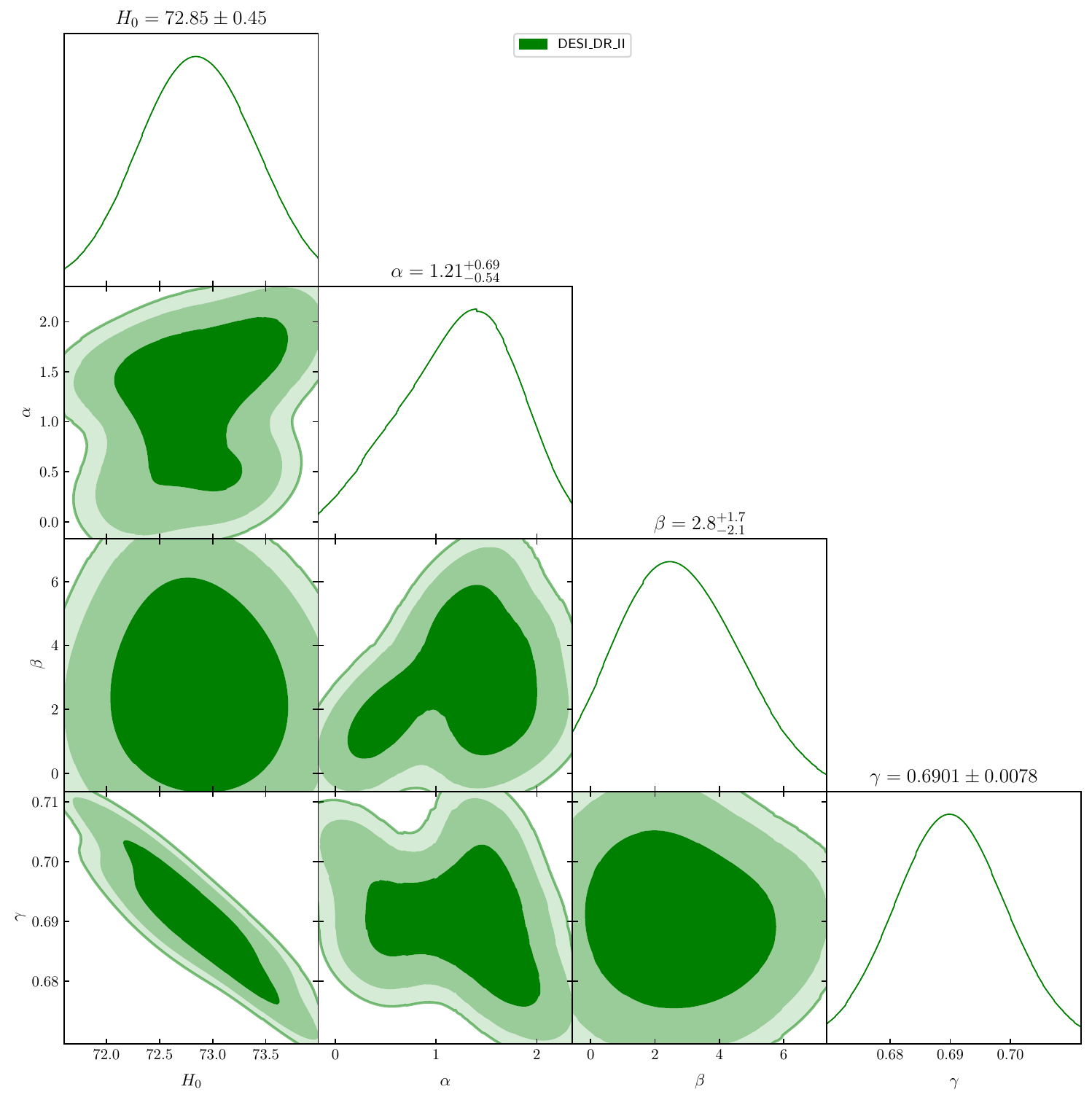}
        \caption{}
    \end{subfigure}
    \begin{subfigure}{0.48\textwidth}
        \includegraphics[width=\textwidth]{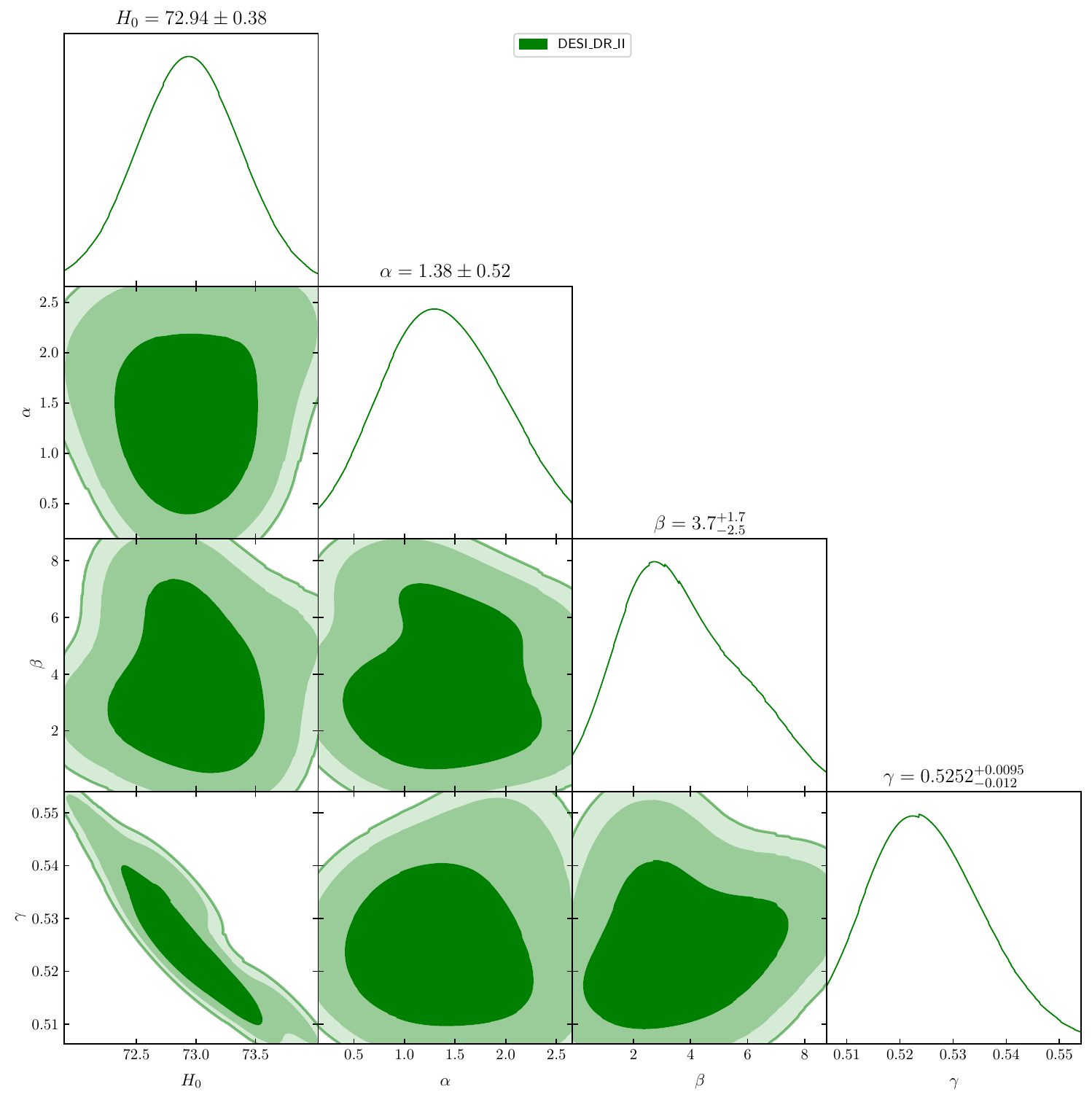}
        \caption{}
    \end{subfigure}
    \caption{Posterior distribution of $f(Q,\mathcal{L}_m)$ gravity for (a)Model-I, (b) Model-II models parameters at $1\sigma$ and $2\sigma$ confidence levels for DESI DR II dataset .}
    \label{desi}
\end{figure*}
\begin{figure*}
    \centering
    \begin{subfigure}{0.48\textwidth}
        \includegraphics[width=\textwidth]{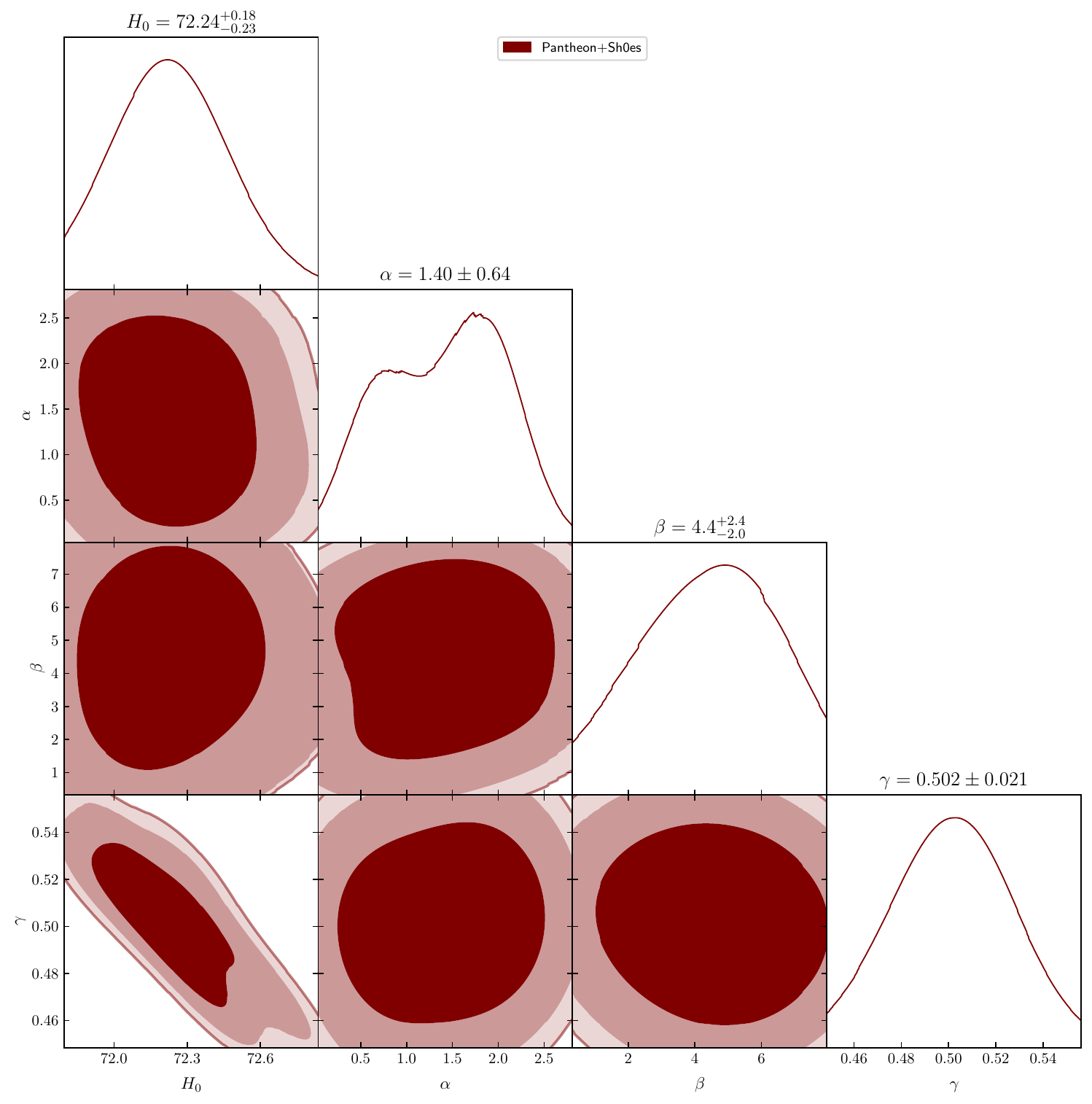}
        \caption{}
    \end{subfigure}
    \begin{subfigure}{0.48\textwidth}
        \includegraphics[width=\textwidth]{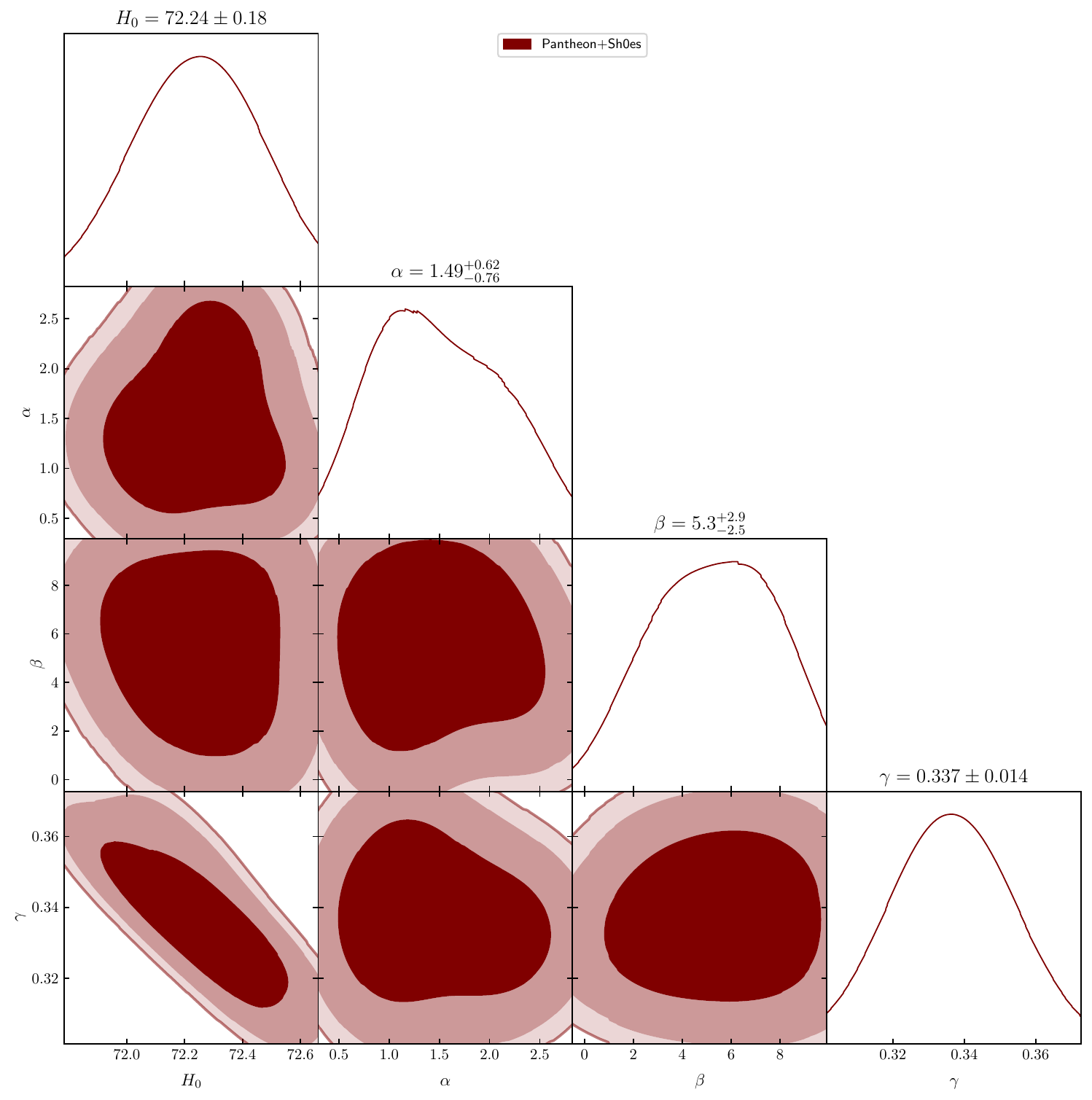}
        \caption{}
    \end{subfigure}
    \caption{Posterior distribution of $f(Q,\mathcal{L}_m)$ gravity for (a)Model-I, (b) Model-II models parameters at $1\sigma$ and $2\sigma$ confidence levels for 1701 $Pantheon^+$ dataset .}
    \label{pan}
\end{figure*}
Table~\ref{tab:model_metrics_comparison} summarizes the AIC, BIC, $\Delta$AIC, and $\Delta$BIC values for Model-I, Model-II, and the reference $\Lambda$CDM model obtained from the CC+BAO, DESI DR2, and Pantheon$^{+}$ datasets. Here, the $\Lambda$CDM model is adopted as the reference model for calculating $\Delta$AIC and $\Delta$BIC, which by definition has $\Delta$AIC$=\Delta$BIC$=0$.

For the CC+BAO dataset, both Model-I and Model-II exhibit relatively large positive values of $\Delta$AIC and $\Delta$BIC, thereby providing strong statistical support in favor of the $\Lambda$CDM model, although Model-II performs marginally better than Model-I. In the Model of the DESI DR2 dataset, the differences are comparatively smaller, suggesting moderate evidence in favor of $\Lambda$CDM while indicating that both model variants remain reasonably competitive. For the Pantheon$^{+}$ dataset, the smallest $\Delta$AIC and $\Delta$BIC values are obtained. Based on the AIC criterion, this result provides positive statistical evidence against the extended models. Overall, the $\Lambda$CDM model remains statistically favored across all datasets according to both the AIC and BIC criteria. Nevertheless, the relatively small information criterion differences obtained for Model-II, particularly for the DESI DR2 and Pantheon$^{+}$ datasets, suggest that it remains a viable alternative and is not decisively ruled out by the current observational data.\\

From Fig.~\ref{cc}, the posterior distributions obtained from the OHD+BAO dataset are unimodal and approximately Gaussian for both Model~1 and Model~2, indicating stable parameter estimation and the absence of multimodal solutions. Among the model parameters, $\gamma$ exhibits the narrowest and most sharply peaked distribution, whereas $\beta$ shows the broadest posterior, reflecting comparatively weaker constraints. The posterior distributions obtained from the DESI DR II dataset are broadly consistent with those of the CC+BAO analysis, exhibiting unimodal one-dimensional distributions and smooth, closed confidence contours, as shown in Fig. \ref{desi}. For the Pantheon$+$SH0ES dataset, the posterior distributions exhibit a similar overall behavior in both Model~I and Model~II, which can be found in Fig.\ref{pan}. The marginalized distribution of $H_0$ is the most sharply constrained, while $\gamma$ also displays a narrow and nearly symmetric posterior. In contrast, the posterior of $\beta$ is comparatively broader and exhibits a mild positive skewness, with the effect being slightly more pronounced in Model~II. The corresponding two-dimensional confidence contours remain well localized and approximately elliptical for all parameter pairs, indicating stable parameter estimation. \\

\begin{figure*}
    \centering
    \begin{subfigure}{0.48\textwidth}
        \includegraphics[width=\textwidth]{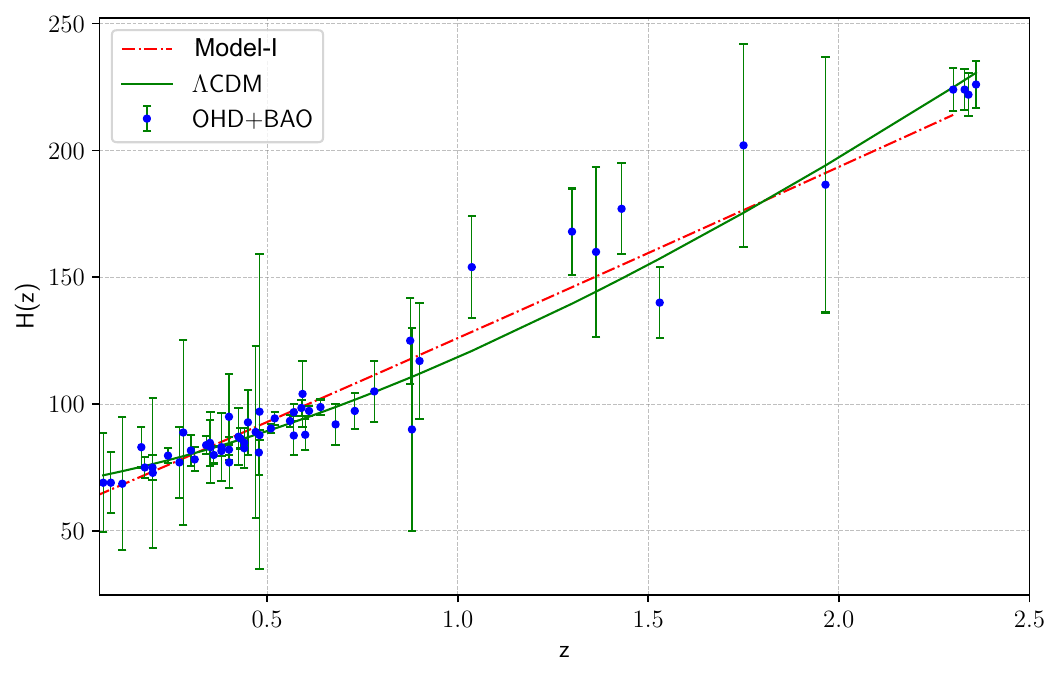}
        \caption{}
    \end{subfigure}
    \begin{subfigure}{0.48\textwidth}
        \includegraphics[width=\textwidth]{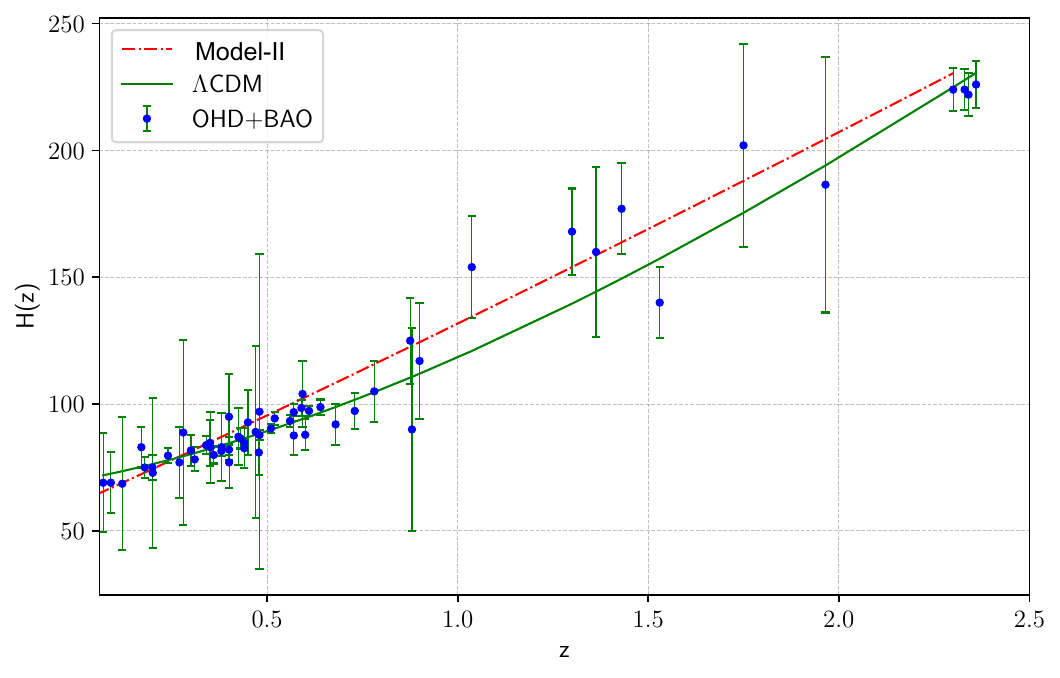}
        \caption{}
    \end{subfigure}
    \caption{Evolution of the Hubble parameter $H(z)$ for (a) Model-I and (b) Model-II reconstructed using the best-fit parameters obtained from the CC+BAO dataset, compared with the standard $\Lambda$CDM model.}
    \label{hubble}
\end{figure*}

Fig. \ref{hubble} presents the reconstructed Hubble parameter, $H(z)$, corresponding to Model-I and Model-II, respectively, obtained using the best-fit parameters from the MCMC analysis of the CC+BAO dataset. The theoretical predictions are compared with the corresponding observational measurements and the standard $\Lambda$CDM model. In both Models, the reconstructed $H(z)$ closely follows the $\Lambda$CDM evolution and agrees well with the observational data across the entire redshift range considered, within the $1\sigma$ confidence limits. Although noticeable deviations appear at intermediate and higher redshifts, in \textbf{Model-I}, the model curve shows relatively close agreement with the $\Lambda$CDM model at low redshift, while some differences become evident towards higher values of $z$. For \textbf{Model-II}, the model exhibits a somewhat larger deviation from the $\Lambda$CDM prediction, particularly at intermediate-to-high redshifts, while still remaining compatible with the observational error bars. These variations can be attributed to the influence of the underlying model parametrization on the expansion history. Nevertheless, the overall agreement between the reconstructed curves and the CC+BAO observations confirms the observational viability of both Models across the investigated redshift range.

%%%%%%%%%%%%%%%%%%%%%%%%%%%%%%%%%%%%%%%%%%%%%%%%%%
%%%%%%%%%%%%%%%%%%%%%%%%%%%%%%%%%%%%%%%%%%%%%%%%
\section{Conclusions}\label{sec7}
%%%%%%%%%%%%%%%%%%%%%%%%%%%%%%%%%%%%

%%%%%%%%%%%%%%%%%%%%%%%%%%%%%%%
In this work, we have investigated two parametrizations of $f(Q,\mathcal{L}_m)$ gravity using CC+BAO, DESI DR2, and Pantheon$^{+}$ data, complemented by a critical-point analysis of the corresponding cosmological dynamics. By constraining the model parameters through an MCMC analysis, we have examined the expansion history, posterior parameter distributions, and statistical performance of the two Models relative to the standard $\Lambda$CDM scenario.

Our results show that the inferred value of the Hubble constant is remarkably insensitive to the choice between the two Models, but strongly dependent on the observational dataset. The CC+BAO combination consistently gives $H_0\simeq68.4~{\rm km\,s^{-1}\,Mpc^{-1}}$, whereas DESI DR2 and Pantheon$^{+}$ favor higher values of approximately $72.9$ and $72.2~{\rm km\,s^{-1}\,Mpc^{-1}}$, respectively. Thus, the $f(Q,\mathcal{L}_m)$ framework accommodates the dataset-dependent preference for $H_0$, but does not by itself eliminate the existing discrepancy between the lower-$H_0$ and higher-$H_0$ probes. This behavior is found to be robust for both parametrizations. The critical-point analysis further demonstrates that the model admits physically relevant cosmological epochs and a stable late-time accelerated attractor, providing a consistent dynamical realization of the observed cosmic expansion history.

The coupling parameters exhibit a stronger dependence on both the observational dataset and the adopted parametrization. In particular, $\gamma$ decreases systematically when moving from CC+BAO to DESI DR2 and subsequently to Pantheon$^{+}$, while the transition from Model-I to Model-II generally produces a lower $\gamma$ and a higher $\beta$. This behavior points to a nontrivial degeneracy between the parameters governing the matter--non-metricity coupling. The broader redshift coverage of CC+BAO provides comparatively tighter constraints on this coupling sector, whereas the predominantly low-redshift Pantheon$^{+}$ sample leaves a more pronounced $\beta$--$\gamma$ correlation. The posterior distributions are nevertheless unimodal and well localized, with approximately elliptical confidence contours, indicating stable and statistically well-behaved parameter estimation without evidence for multimodal solutions.

The reconstructed expansion histories provide an additional consistency test of the model. For both Models, $H(z)$ remains in good agreement with the CC+BAO observations and closely tracks the $\Lambda$CDM prediction over the redshift range investigated. Model-I generally remains closer to the $\Lambda$CDM expansion history, particularly at low redshift, while Model-II allows somewhat larger deviations at intermediate and higher redshifts. Importantly, these deviations remain compatible with the observational uncertainties, demonstrating that both parametrizations can reproduce the observed background expansion history without introducing significant tension with current late-time data.

The information-criterion analysis, however, provides a more conservative assessment of the models. The $\Lambda$CDM model exhibits a stronger statistical preference than both
 Model-I and Model-II for all three datasets according to both AIC and BIC. The preference is strongest for CC+BAO, while the differences become considerably smaller for DESI DR2 and Pantheon$^{+}$. In particular, Model-II performs somewhat better than Model-I for CC+BAO, although both remain disfavored relative to $\Lambda$CDM. For DESI DR2 and Pantheon$^{+}$, the relatively modest information-criterion differences indicate that the extended $f(Q,\mathcal{L}_m)$ scenarios cannot be decisively excluded by the present data despite the statistical advantage of $\Lambda$CDM.

Overall, the results of our analysis suggest that the proposed $f(Q,\mathcal{L}_m)$ parametrizations provide observationally consistent extensions to the conventional cosmological model, particularly in describing the background evolution of the Universe.
 They successfully reproduce the observed $H(z)$ evolution and remain compatible with current late-time measurements, while allowing nontrivial constraints on the coupling sector. At the same time, the information criteria show that the additional model freedom is not presently required by the data. The systematic variation of $\gamma$ across datasets and its correlation with $\beta$ further suggest that the coupling sector deserves investigation beyond the homogeneous background level. Future high-precision BAO measurements, Type Ia supernovae, cosmic chronometers, and especially independent growth-of-structure observations such as $f\sigma_8$, together with higher-redshift probes, will be crucial for breaking the parameter degeneracies and determining whether the departures from $\Lambda$CDM suggested by $f(Q,\mathcal{L}_m)$ gravity have a genuine observational signature.\\

%%%%%%%%%%%%%%%%%%%%%%%%%%%%%%%%%%%%%%%%%%%%%%%%%
\textbf{Acknowledgments :} RM acknowledges UGC, Govt. of India, for providing Senior Research Fellowship (NTA Ref. No.: 211610083890) and AK is thankful to IIEST, Shibpur, India, for providing Institute Fellowship (SRF).
 
\bibliographystyle{naturemag}
\bibliography{ref.bib}

\end{document}